\documentclass{pasj00}
\SetRunningHead{Astronomical Society of Japan}{Usage of \texttt{pasj00.cls}}
\RelativeFigurePath{./}
\usepackage{times}
\begin{document}
\SetRunningHead{T. Sato et al.}{Suzaku Observations of the Hydra A Cluster}

\title{Suzaku observations of the Hydra A cluster out to the virial radius}
\author{
Takuya \textsc{Sato}\altaffilmark{1}, 
Toru \textsc{Sasaki}\altaffilmark{1}, 
Kyoko \textsc{Matsushita}\altaffilmark{1}, 
Eri \textsc{Sakuma}\altaffilmark{1}, 
Kosuke \textsc{Sato}\altaffilmark{1},\\
Yutaka \textsc{Fujita}\altaffilmark{2},
Nobuhiro \textsc{Okabe}\altaffilmark{3}, 
Yasushi \textsc{Fukazawa}\altaffilmark{4},
Kazuya \textsc{Ichikawa}\altaffilmark{1},\\
Madoka \textsc{Kawaharada}\altaffilmark{5},
Kazuhiro \textsc{Nakazawa}\altaffilmark{6}, 
Takaya \textsc{Ohashi}\altaffilmark{7},
Naomi \textsc{Ota}\altaffilmark{8},\\
Motokazu \textsc{Takizawa}\altaffilmark{9},
and
Takayuki \textsc{Tamura}\altaffilmark{5} 
}
\altaffiltext{1}{Department of physics, Tokyo University of Science, 1-3 Kagurazaka, Shinjuku-ku, Tokyo 162-8601 , Japan}
\email{j1209703@ed.kagu.tus.ac.jp; matusita@rs.kagu.tus.ac.jp}
\altaffiltext{2}{Department of Earth and Space Science, Graduate School of Science, Osaka University, 
Toyonaka, Osaka 560-0043, Japan}
\altaffiltext{3}{Institute of Astronomy and Astrophysics, Academia Sinica, P.O. Box 23-141, Taipei 106, Taiwan; Astronomical Institute, Tohoku University, Aramaki, Aoba-ku, Sendai, 980-8578, Japan}
\altaffiltext{4}{Department of Physical Science, Hiroshima University, 1-3-1 Kagamiyama, Higashi-Hiroshima, Hiroshima 739-8526, Japan}
\altaffiltext{5}{Institute of Space and Astronautical Science, Japan Aerospace Exploration Agency, 3-1-1 Yoshinodai, Chuo-ku, Sagamihara, Kanagawa 252-5210, Japan}
\altaffiltext{6}{Department of Physics, The University of Tokyo, 7-3-1 Hongo, Bunkyo-ku, Tokyo 113-0033, Japan}
\altaffiltext{7}{Department of Physics, Tokyo Metropolitan University, 1-1 Minami-Osawa, Hachioji, Tokyo 192-0397, Japan}
\altaffiltext{8}{Department of Physics, Nara Women's University, Kitauoyanishi-machi, Nara, Nara 630-8506, Japan}
\altaffiltext{9}{Department of Physics, Yamagata University, Yamagata, Yamagata 990-8560, Japan}

\KeyWords{
galaxies: clusters: individual (Hydra~A cluster) --- X-rays: galaxies: clusters --- intergalactic medium
}
\Received{---}
\Accepted{---}
\Published{---}

\maketitle

\begin{abstract}
We report Suzaku observations of the northern half of the
 Hydra A cluster out to $\sim$ 1.4 Mpc, reaching  the virial radius.
This is the first Suzaku observations of a medium-size ($kT\sim $3 keV) cluster
out to the virial radius.
Two observations were
conducted, north-west and north-east offsets,
which continue in a filament direction and a void direction of
 the large-scale structure of the Universe, respectively. 
The X-ray emission and distribution of galaxies  elongate in the filament direction.
The temperature profiles in 
 the two directions are mostly consistent with each other
within the error bars and drop to 1.5 keV
at 1.5 $r_{500}$.
As observed by Suzaku in hot clusters,
the entropy profile becomes flatter beyond $r_{500}$, in disagreement
 with the $r^{1.1}$ relationship
 that is expected from accretion shock heating models.
When scaled with the average intracluster medium (ICM) temperature, 
the entropy profiles of clusters observed with Suzaku are universal and do not
depend  on system mass.
The hydrostatic mass values in the void and  filament directions are in good 
agreement, 
and the Navarro, Frenk, and White  universal 
mass profile represents the  hydrostatic mass 
distribution up to $\sim 2~r_{500}$. 
Beyond $r_{500}$, the ratio of gas mass to hydrostatic mass exceeds the
result of the Wilkinson microwave anisotropy probe,
 and at $r_{100}$, these ratios in the filament and void directions
reach 0.4 and 0.3, respectively.  
We discuss  possible deviations from hydrostatic equilibrium at cluster outskirts.
We derived radial profiles of
the gas-mass-to-light ratio and
 iron-mass-to-light ratio out to the virial radius.
 Within $r_{500}$, the iron-mass-to-light ratio
of the Hydra A cluster  was compared with those in other clusters observed with Suzaku. 
\end{abstract}

\section{Introduction}

Clusters of galaxies are the largest self-gravitating systems
in the Universe,
and offer unique information on the process of
structure formation governed by cold dark matter (CDM)\@.
In addition, these clusters are considered as a laboratory for 
studying thermal and chemical evolutions of the Universe in which 
baryons play the most important role. 
X-ray observations provide valuable information about the structure
formation, gas heating and cooling, and metal enrichment of galaxy clusters.
Because the dynamical time-scale of clusters
is comparable to the Hubble time, cluster outskirts should maintain
original records of cluster evolution via accretion of gas and 
substructures from filaments of the surrounding large-scale structure of 
the Universe.

Thanks to the low and stable background of the X-ray Imaging 
Spectrometer (XIS; \cite{koyama07}),
 Suzaku \citep{mitsuda07} was able to
unveil for the first time the intracluster medium (ICM) 
beyond $r_{500}$, a radius within which the mean cluster-mass density is 500
times the cosmic critical density.
The accurate calibration 
of the XIS also allows precise measurements of the ICM temperature \citep{Sato2011}.
Suzaku derived the temperature and entropy profiles of the ICM of several 
massive clusters up to the virial radius
(\cite{george09}; \cite{reiprich09}; \cite{bautz09}; \cite{kawaharada10}; \cite{hoshino10}; 
\cite{simionescu11}; \cite{akamatsu11}).
From the center to $r_{200}$, 
a systematic drop in temperature was found by a factor of $\sim$ 3,
 and the observed entropy profiles become flatter beyond $r_{500}$.
These profiles are lower than the  entropy profile predicted by
 the numerical simulations 
of gravitational collapse (\cite{tozzi01}; \cite{voit05}), 
which is proportional to $r^{1.1}$.
 \citet{kawaharada10} discovered that  beyond $r_{500}$ of the Abell 1689 cluster, 
the total mass obtained from weak-lensing observations with Subaru
is larger than that calculated  assuming hydrostatic equilibrium.
Therefore, one explanation for the low  entropy profiles
at cluster outskirts is that infalling matter  retained some of its kinetic
energy in bulk motion \citep{bautz09, george09, kawaharada10}. 
Based on Suzaku observations of the Perseus cluster, 
\citet{simionescu11} proposed a gas-clumping effect  as an additional interpretation.
 \citet{hoshino10} and \citet{akamatsu11}
discussed possible deviations of electron
temperature from ion temperature to explain the observed lower
temperature and entropy profiles.
To clarify these effects in the outskirts, 
dependence on the system mass or on the ICM temperature should be examined.
With XMM-Newton observations,
\citet{urban11}  found a  similar 
flattening of the entropy profile in one direction  in the  Virgo cluster ($kT$ = 2.3~keV).
However, \citet{humphrey11} detected no evidence of flat profiles at
large scales ($> r_{500}$) in the relaxed fossil group or in the poor cluster, RXJ 1159+5531.
Therefore, more samples of medium-sized clusters are required.

In addition to measuring the temperature and entropy profiles, 
Suzaku  measured the abundance of Fe in the ICM 
beyond 0.5$r_{180}$ \citep{fujita08,tawa08, simionescu11}.
Metal abundances in the ICM also provide important information
on the chemical history and evolution of clusters.
The ASCA satellite first measured the distribution of Fe in the ICM
(\cite{fukazawa00}; \cite{finoguenov01}).
Recently, spatial distributions of Fe from 0.3 to 0.4 $r_{180}$ have been studied
with the XMM-Newton and Chandra satellites (\cite{vikhlinin05}; 
\cite{maughan08}; \cite{leccardi08}; \cite{matsushita11}).
Since metals are synthesized by supernovae (SNe) in
galaxies, the ratios of metal mass in the ICM to the total light
from galaxies in clusters or groups, (i.e., metal-mass-to-light ratios)
are key parameters in investigating the chemical evolution of the
ICM\@.
Suzaku measured the iron-mass-to-light ratios (IMLR) of several clusters 
and galaxy groups out to  0.2 $\sim$ 0.5 $r_{180}$
(\cite{matsushita07}; \cite{komiyama09}; \cite{sato07}; \cite{sato08}; 
\cite{sato09a}; \cite{sato09b}; \cite{sato10}; \cite{sakuma11}),
and with XMM-Newton, the IMLR of the Coma cluster out to 0.5$r_{180}$ was derived
\citep{matsushita2011b}.
The IMLR profiles increase with radius,
indicating that Fe in the ICM extends farther than stars. 

The Hydra A cluster (z = 0.0539) with an ICM temperature of $\sim$3 keV
is one of 
the prototype cool-core clusters in which  \citet{mcnamara00} discovered
 a displacement of X-ray gas in the central region through the radio lobes
from the central active galactic nucleus (AGN).
This cluster is also known as Abell 780 and remains a
major examples of AGN interaction that is 
studied through radio and X-rays \citep{taylor1990,david01,nulsen02,nulsen05,
lane04,wise07,simonescu09a,simonescu09b,kirkpatrick09}.
Through Chandra and XMM-Newton observations, 
a sharp X-ray surface-brightness edge was detected at radii between
$4.3'$ and $6'$ (200--300 kpc) and was interpreted as a shock wave caused by an AGN outburst.
The abundances of the Hydra A cluster was measured within $\sim$0.3 $r_{180}$
with XMM \citep{simonescu09b,matsushita11}.

This paper reports the results of two Suzaku observations in the
northern half of the  Hydra A 
cluster out to 25$'$($\simeq$1.5 Mpc), which corresponds to the virial radius.
The observations were conducted using the XIS.
This study reports the first Suzaku observations of a medium-sized cluster with 
an average temperature of $\sim $3 keV up to the virial radius.
The two observed fields continue in 
 a filament direction that continues to the Abell 754 cluster
 and a void direction
 of the large-scale structure of the Universe (see Figure \ref{fig:extended source}).

In this study, we use the Hubble constant, $H_{\rm 0} = 70$~km~s$^{-1}$~Mpc$^{-1}$\@,
$\Omega_{m}=0.27$ and $\Omega_{\Lambda}=0.73$.
The luminosity distance $D_{\rm L}$ and angular size distance $D_{\rm A}$ 
to the Hydra~A cluster are $D_{\rm L}=240$~Mpc and $D_{\rm A}=217$~Mpc, respectively,
and $1'$ corresponds to 66.6~kpc. 
We use the epoch J2000.0 for the right ascension and declination of the equatorial
coordinate system.
In addition, we use the solar abundance table by \citet{lod03},
in which the solar Fe abundance relative to H is 2.95$\times$10$^{-5}$.
Considering a difference in solar He abundance, 
the Fe abundance yielded by \citet{lod03}  
is 1.5 times higher than that using the photo-spheric value by \citet{angr}.
Errors were quoted at a 90\% confidence level for a single parameter.

\section{Observation and Data Reduction}
\label{sec:obs}

\subsection{Suzaku observations}
\label{sec:suzakuobs}

\begin{table*}[th]
\caption{Observation log of the Hydra A cluster}
\label{tab:ob}
\begin{tabular}{ccccccc} 
\hline
Field name & Target name & Sequence & Observation & RA\footnotemark[$*$] & Dec\footnotemark[$*$] & Exposure\\ 
 & & number & date & (deg) & (deg) & (ks) \\
\hline
North-west (filament) & HYDRA A-1 & 805007010 & 2010-11-08 & 139.3723 & -11.9472 & 35.6 \\
North-east (void) & HYDRA A-2 & 805008010 & 2010-11-09 & 139.6749 & -11.9472 & 34.9 \\
\hline
\multicolumn{7}{@{}l@{}}{\hbox to 0pt{\parbox{170mm}{\footnotesize
\footnotemark[$*$] 
Coordinates are referred to J2000.0.
\par\noindent
 }\hss}}
\end{tabular}
\end{table*}

\begin{figure}[t]
\begin{center}
\FigureFile(80mm, 60mm){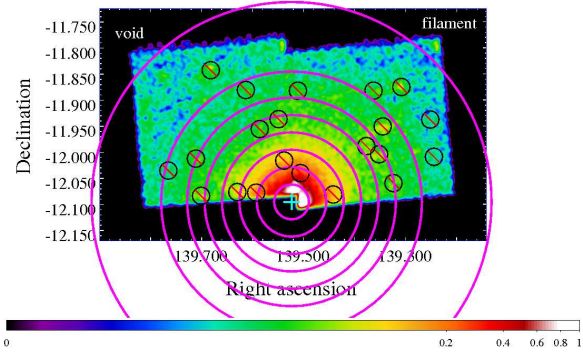}
\end{center}
\caption{XIS0 image of the Hydra~A cluster within 0.5--4.0 keV energy range.
The light blue cross indicates the X-ray peak of the cluster.
Differences in the vignetting effect are uncorrected.
Regions of spectral accumulation are shown as magenta rings.
The ring radii are $2'$, $4'$, $6'$, $8'$, $10'$, $12'$, $15'$ and $23'$.
Black circles represent excluded regions around point sources.}
\label{fig:image}
\end{figure}

Suzaku conducted two observations of the Hydra~A cluster in November 2010, which 
was during the Suzaku  Phase-V period.
The details of the observations are summarized in Table \ref{tab:ob}. 
The first observation, HYDRA A-1, was
 $12'$ north-west offset from the X-ray peak of the Hydra~A cluster
with coordinates (RA, Dec) = (139.5236, -12.0955) in degrees.
The second observation, HYDRA A-2, was
$12'$ north-east offset from the X-ray peak of the Hydra~A cluster. 
The north-west and north-east fields
continue into the filament and void structures, respectively 
(See figure \ref{fig:extended source}).
Hereafter, we refer to the north-west and north-east offsets as filament
and void, respectively.

The XIS, which was operated in its normal mode during the observations, 
consists of three sets of X-ray CCD 
(XIS0, XIS1, and XIS3). XIS1 is a back-illuminated (BI) sensor, while both
XIS0 and XIS3 are front-illuminated (FI) sensors.
Data reduction was done with HEAsoft version 6.11\@. 
The XIS event lists created by  rev~2.5 pipeline processing 
were filtered using the following additional criteria: a geomagnetic cutoff 
rigidity (COR2) $>$6 GV, and an elevation angle $>10^{\circ}$ from the earth limb.
The 5$\times$5 and 3$\times$3 editing modes data formats
were added. The exposure times after data selection are shown 
in Table \ref{tab:ob}. 

To obtain the temperature and electron density profiles,
we accumulated spectra within annular regions centered
on the X-ray peak as shown in Figure \ref{fig:image}.
Regions around calibration sources and
 the innermost region within 2$'$ were excluded from the spectral analysis.
The reason for this exclusion is that we could not obtain sufficient photon
statistics to analyze the spectra (which have multiphase features at the cluster
center) because the given area was detected on the edge of a CCD chip
in our observation.
 In addition, Suzaku's PSF is also insufficient to
resolve such a complex structure. 
However, with their powerful imaging capability,
 Chandra and XMM observations  have unveiled such complex features in the central region.
Using the ewavelet tool in SAS\footnote{http://xmm.esa.int/sas/}, 
we searched for point-like sources in the Suzaku images in the 
energy range of 2.0--10.0 keV\@.  In this energy range, the flux levels 
were approximately $> 2~\times 10^{-14}$ erg cm$^{-2}$ s$^{-1}$.
These  point sources were excluded as circular regions with radii of 1$'$ (figure \ref{fig:image}).
The non X-ray background (NXB) was subtracted 
from each spectrum using a database of night Earth observations with 
the same detector area and COR distribution \citep{tawa08}.

We included the degradation of energy resolution due to 
radiation damage in the redistribution matrix file (RMF) generated by 
the {\tt xisrmfgen} Ftools task. In addition, we created an ancillary response 
file (ARF) using the {\tt xissimarfgen} Ftools task \citep{ishisaki07}. 
A decrease in the low-energy transmission of the XIS optical blocking 
filter (OBF) was also included in the ARF. For filament and void fields,
we generated ARF files assuming uniformly extended emission from an enriched
region with a 20$'$ radius.  
We used the XSPEC\_v12.7.0 package and ATOMDB\_v2.0.1 for spectral analysis.
Each spectrum was binned and each spectral bin contained a minimum of 50 counts.
To avoid systematic uncertainties in the background,
we ignored energy ranges above 7 keV and below 0.7 keV. 
In addition, we excluded the narrow energy 
band between 1.82 and 1.84 keV in the fits because of the incomplete 
response around the Si edge.

\subsection{XMM observations}
\label{sec:xmmobs}

We used the same data and analysis as that reported by \citet{matsushita11}.
The XMM-Newton archival data of Hydra A (observation identifier 0109980301)
had exposure times of 14.0 ks, 19.1 ks and 22.2 ks for MOS1, MOS2 and PN, respectively, 
after background flares were screened out.
We selected events with patterns
smaller than 5 and 13 for the PN and MOS, respectively.
Spectra were accumulated in concentric annular regions of $0'$-$0.5'$, $0.5'$-$1'$, 
$1'$-$1.5'$, $1.5'$-$2'$, $2'$-$3'$, $3'$-$4'$,  $4'$-$5'$, $5'$-$6'$, $6'$-$7'$, 
$7'$-$8'$, and $8'$-$9'$,
centered on the X-ray peak of the Hydra A cluster.
Here, the X-ray peaks were derived using the ewavelet
tool of SAS-v8.0.0, and luminous point sources were excluded.
The spectra from MOS1 and MOS2 were added.
The background spectrum for each annular region 
was calculated by integrating blank-sky data in the same detector region.
From deep-sky observations
with XMM-Newton, we selected the data having  background most
similar to that of the Hydra A cluster and the faintest Galactic emission,
after screening out background flare events from the data and the
background, following \citet{katayama04}.
Next, we scaled the background
 using a count rate between 10 and 12 keV.

The response matrix file and the ARF corresponding to 
each spectrum were calculated using SAS-v8.0.0. 
Further details appear in sections 2 and 3 of \citet{matsushita11}.

\section{Data Analysis}
\label{sec:analysis}

\subsection{Estimation of background spectra}
\label{subsec:bkg}

\begin{table*}[tb]
\caption{Resultant parameters of the background components. 
}
\label{tab:bkgtable}
\begin{tabular}{cccc} 
\hline
Cosmic X-ray Background & Local Hot Bubble & Milky-Way halo & \\
\hline \hline
normalization\footnotemark[$\ast$] & normalization\footnotemark[$\dagger$] & normalization\footnotemark[$\dagger$] & \multicolumn{1}{c}{Reduced-$\chi^{2}$}  \\
 & $\times 10^{-3}$ & $\times 10^{-4}$ & $\chi^2$/d.o.f \\
\hline 
$8.92^{+0.72}_{-0.85}$ & $8.93^{+6.02}_{-5.75}$ & $1.83^{+1.85}_{-1.72}$ & 1.04(144/139)\\
\hline
\multicolumn{4}{@{}l@{}}{\hbox to 0pt{\parbox{170mm}{\footnotesize
\footnotemark[$\ast$] 
Measured at 1keV with unit of photons cm$^{-2}$ s$^{-1}$ keV$^{-1}$ sr$^{-1}$.
\par\noindent
\footnotemark[$\dagger$] Normalization of the $APEC$, component divided by the solid angle, 
$\Omega^{\emissiontype{U}}$, 
assumed in the uniform-sky ARF calculation (\timeform{20'} radius), 
$Norm = \int n_{e}n_{H}dV/(4\pi(1+z)^{2}D_{A}^{2})/\Omega^{\emissiontype{U}}\times 10^{-14}$
~cm$^{-5}$~400$\pi$~arcmin$^{-2}$, where $D_{A}$ is the angular distance to the source.
\par\noindent
 }\hss}}
\end{tabular}
\end{table*}

To derive the radial profiles of the  temperature, electron density, and Fe abundance of the
ICM,  we fitted the  NXB-subtracted spectra with a thermal plasma model
(APEC: \cite{smith01}).

We first fitted the spectra in the outermost ring (15$'$--23$'$ region) 
to determine the local X-ray background.
As reported by \citet{Yoshino2009},
the background emission of Suzaku XIS
can be fitted with a three-component model: 
two thermal plasma models (APEC; \cite{smith01}) for the local 
hot bubble (LHB) and the solar wind charge exchange (SWCX), 
the Milky Way Halo (MWH), and a power-law model for 
the extragalactic cosmic X-ray background (CXB). MWH and CXB 
components were convolved with an photoelectric absorption in the Galaxy, $N_{\rm H}$.
Therefore, we adopted the  model formula,
$apec_{\rm LHB}+wabs\times(apec_{\rm MWH}+power$-$law_{\rm CXB})$\@ as  background.
We assumed a zero redshift and a solar abundance for LHB and 
MWH components. The temperature of the
 LHB and MWH was fixed at 0.1 keV and 0.3 keV, respectively. 
The column density of the Galactic neutral hydrogen was fixed at
4.7$\times 10^{20}$ cm$^{-2}$ \citep{kalberla05}, and
the photon index of the CXB component was fixed at 1.41 \citep{kushino02}.
Using the formula
$apec_{\rm LHB}+wabs\times (apec_{\rm ICM}+apec_{\rm MWH}+power$-$law_{\rm CXB})$,
we fitted the spectra with a sum of the background components and a
single-temperature APEC model for the ICM.
Temperature and normalization of the ICM component were allowed to vary, 
and redshift was fixed at 0.0539.
Normalizations of the background components were also left free.
The spectra of the filament and void fields were fitted simultaneously,
and each parameter of background and abundance of the ICM 
was assumed to be the same.

\begin{figure}[th]
\FigureFile(80mm, 60mm){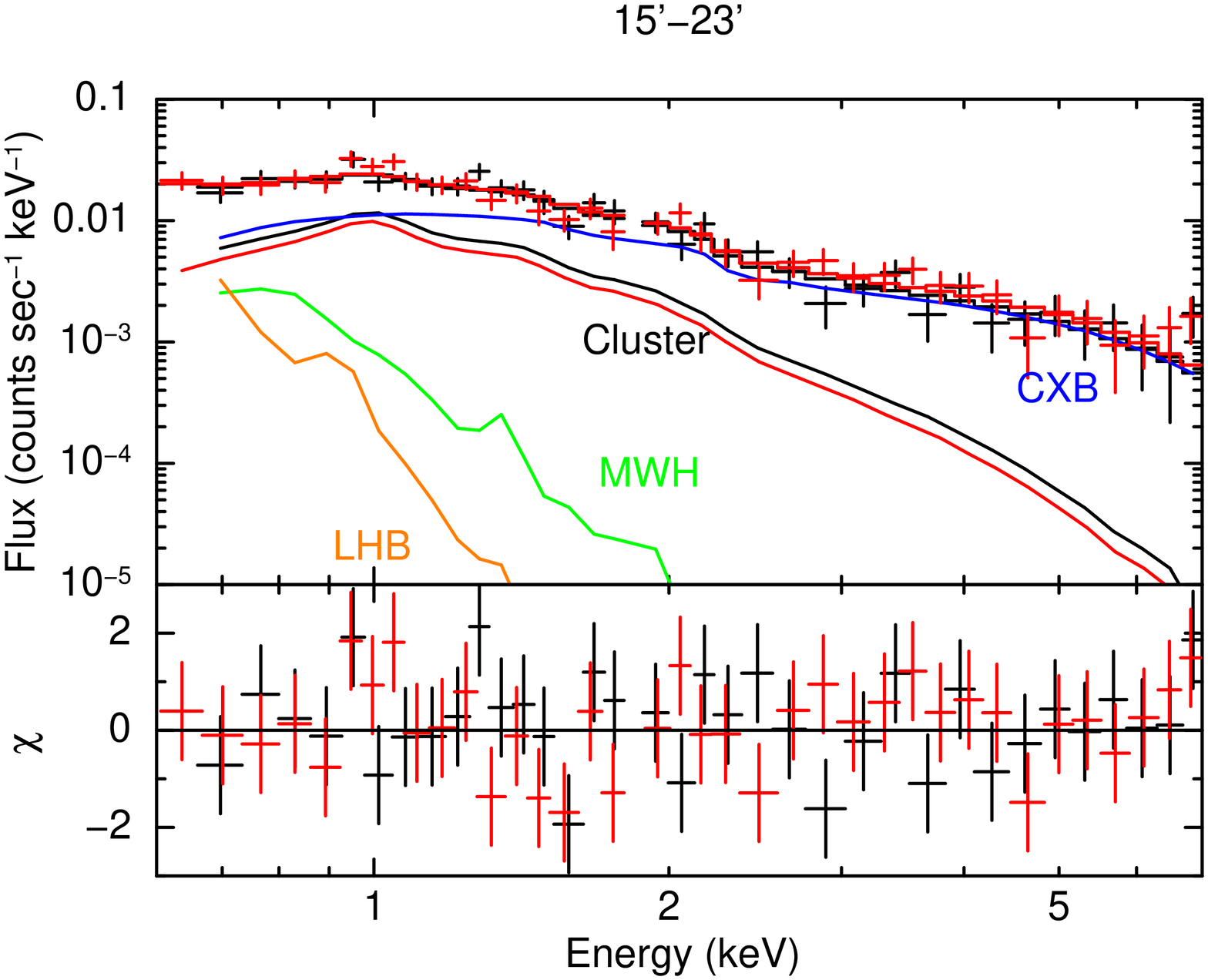}
\caption{NXB--subtracted spectra of XIS1 (crosses) 
for the outermost ring (15$'$--23$'$ region), 
fitted by the $apec_{\rm LHB}+wabs\times (apec_{\rm ICM}+apec_{\rm MWH}+power$-$law_{\rm CXB})$ model
(stepped solid lines).
Black and red colors correspond to the directions in
 the filament and void, respectively. 
The bottom panel shows  residuals of the fit.
The contributions of the ICM are plotted as black and red solid lines,
 those of the Galactic emission  as orange and green lines,
and those of the CXB as blue lines.
}
\label{fig:bkgfit}
\end{figure}

Results of the spectral fit are shown in Figure \ref{fig:bkgfit}.
The resulting parameters for the background and  
ICM are shown in Table \ref{tab:bkgtable} and \ref{tab:results},
respectively.
The  normalizations obtained for the CXB and the Galactic components 
are consistent with 
those derived by \citet{kushino02} and \citet{Yoshino2009}, respectively.

Several blank fields observed with Suzaku contained emission with $kT=0.6-0.8$ keV
\citep{Yoshino2009}.
Therefore, we added an APEC component with $kT=0.6$ keV and refitted the outermost spectra. 
The normalization of the 0.6 keV component had a much smaller value than 
the other Galactic components, and the
 temperatures and normalizations of the ICM  did not change.

\begin{figure*}[thbp]
\FigureFile(80mm, 60mm){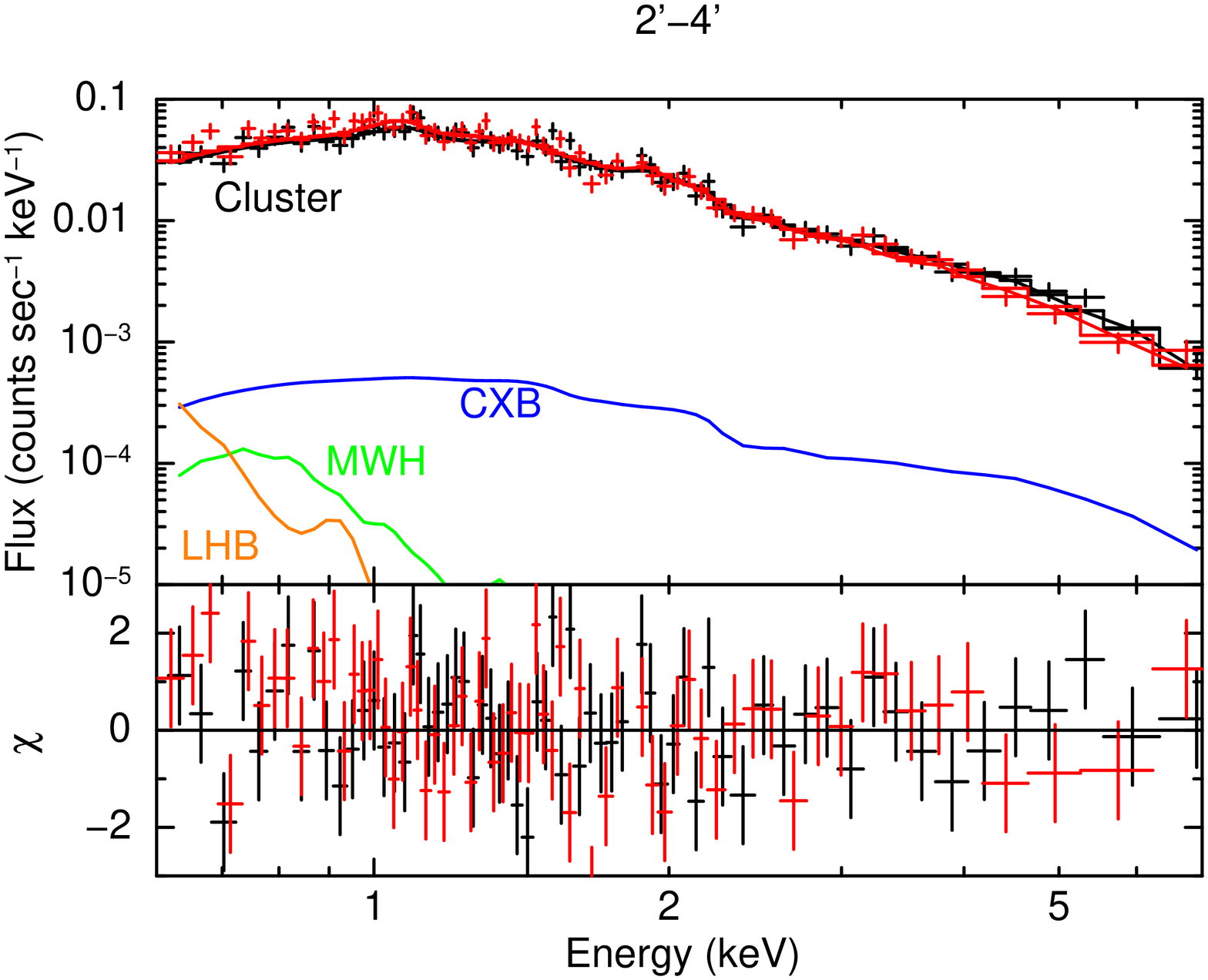}
\FigureFile(80mm, 60mm){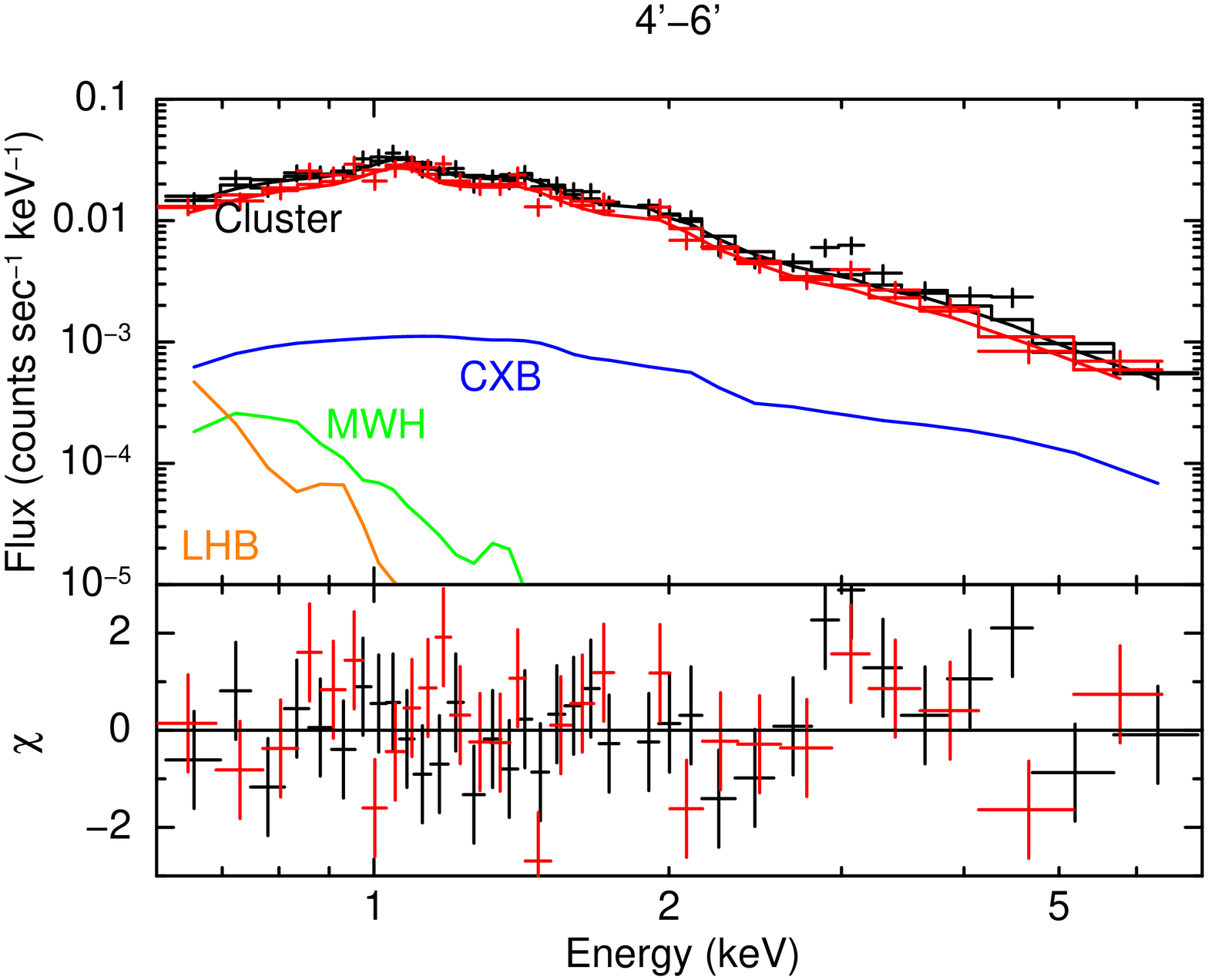}

\FigureFile(80mm, 60mm){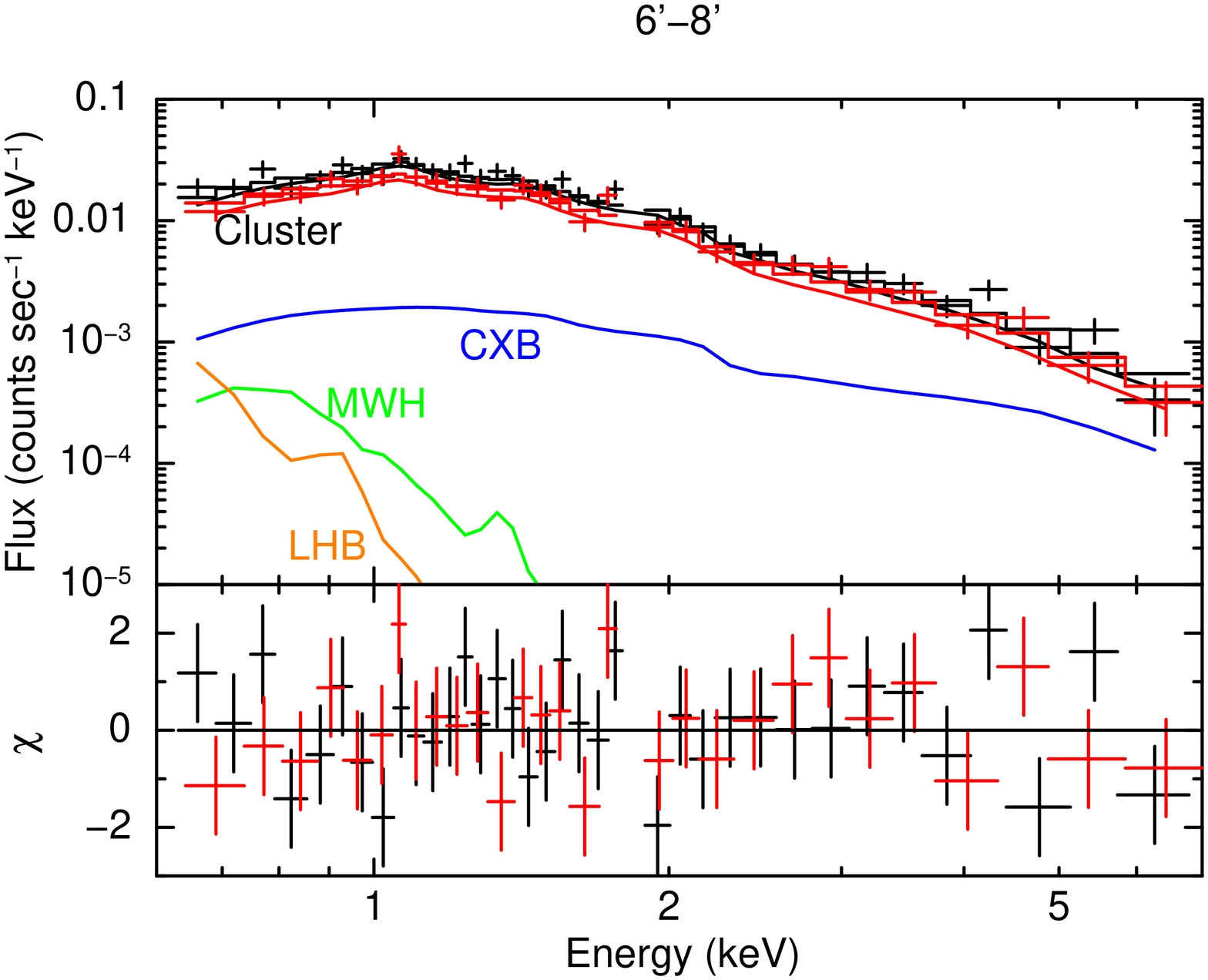}
\FigureFile(80mm, 60mm){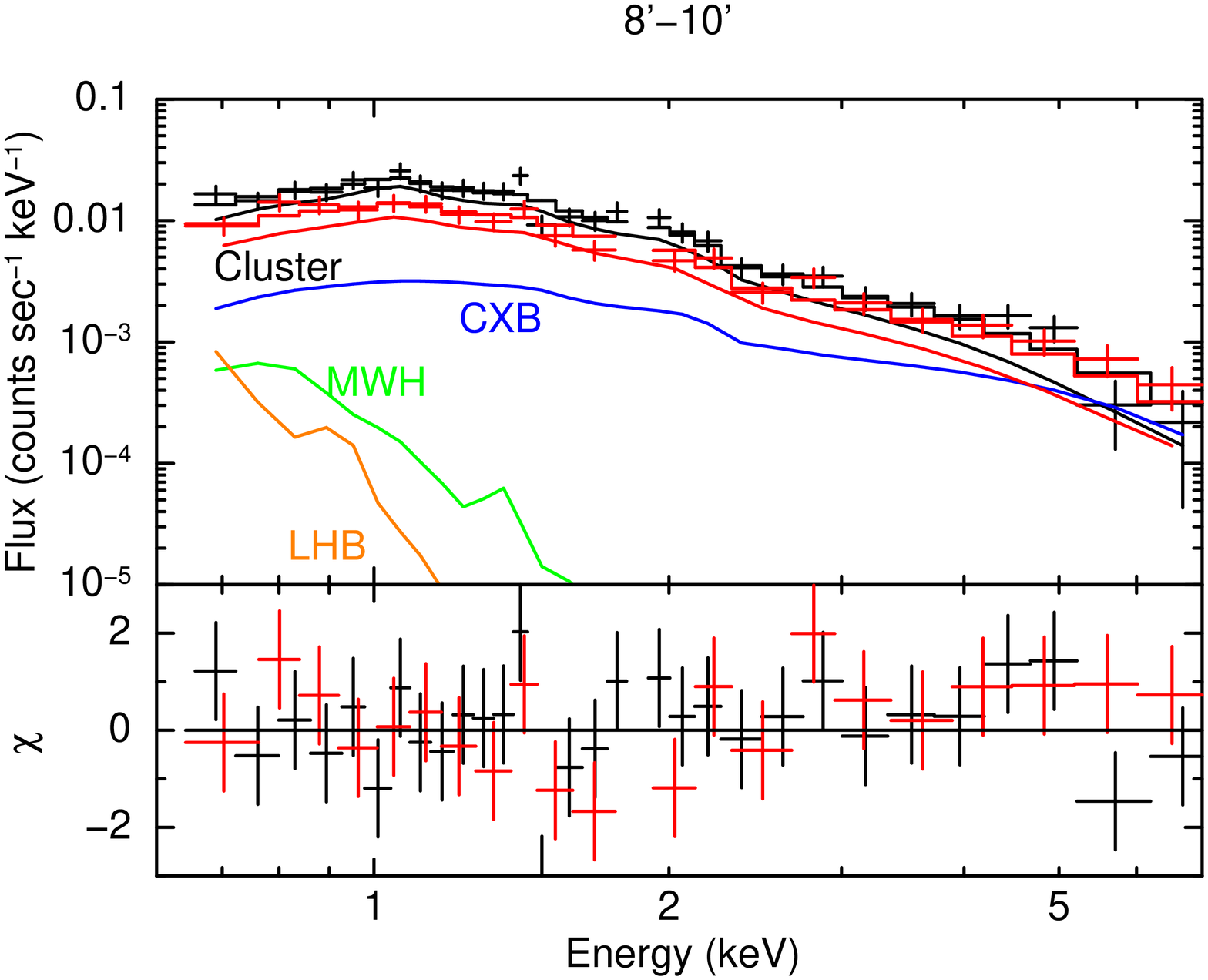}

\FigureFile(80mm, 60mm){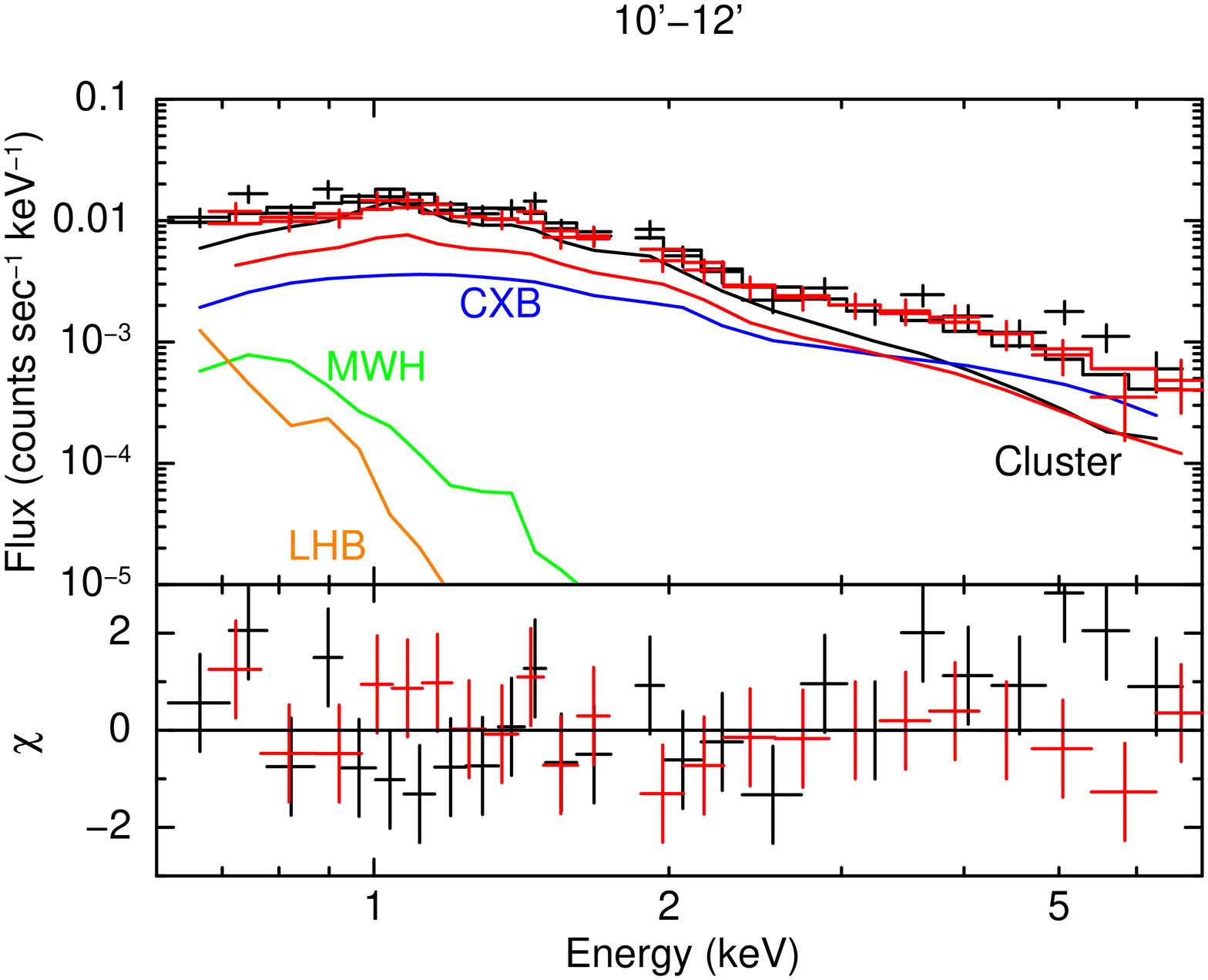}
\FigureFile(80mm, 60mm){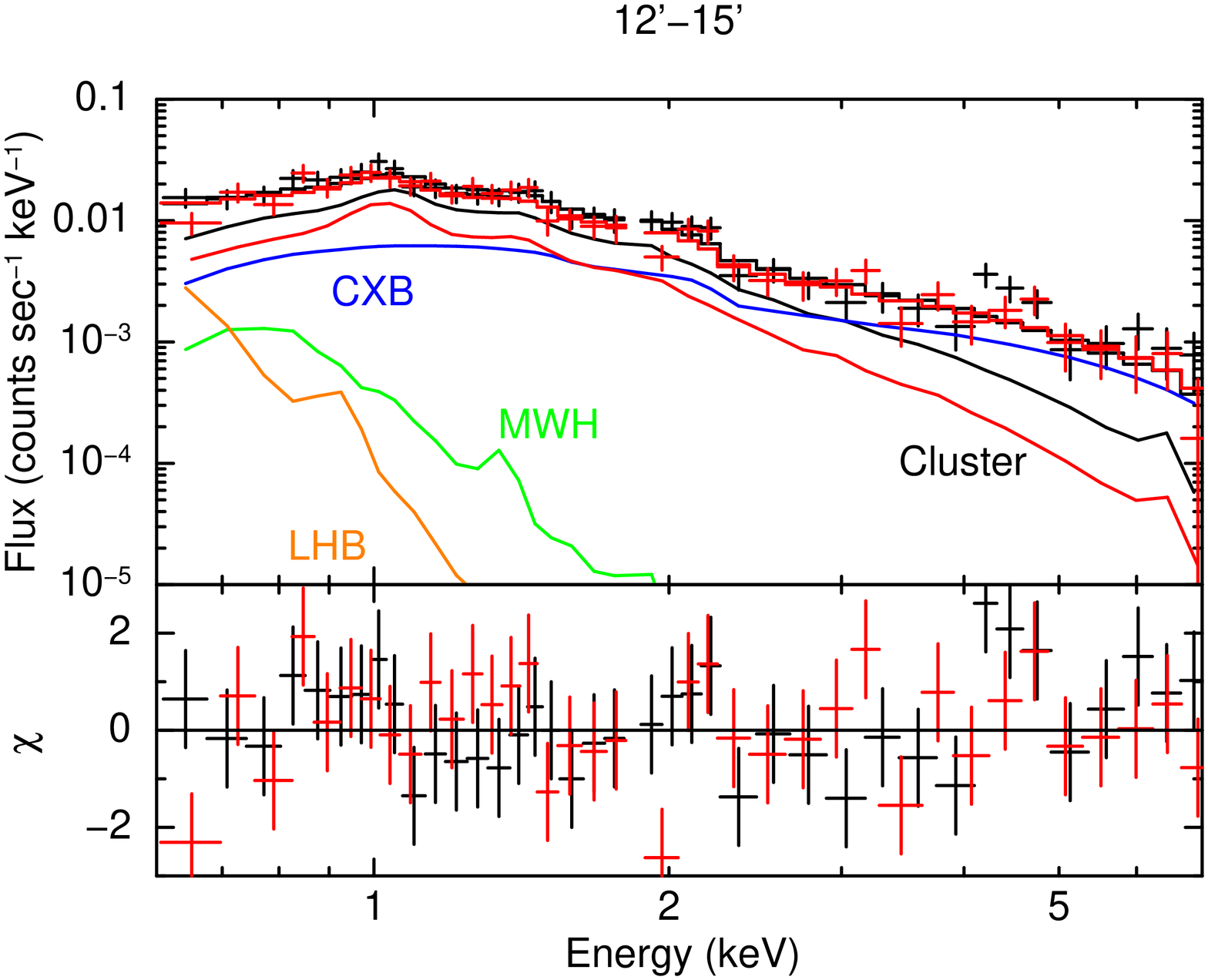}

\caption{NXB-subtracted XIS1 spectra 
fitted with  the single temperature (APEC) model for the ICM.
Black and red colors correspond to directions toward the filament and void, 
respectively. Lower panels in each diagram show residual of fit. 
The contributions of the ICM are plotted as red and black solid lines, 
while those of the galactic emission and the CXB are plotted as orange, green and blue lines,
respectively.
}
\label{fig:spec}
\end{figure*}
\subsection{Spectral fits of annular regions}
\label{subsec:single}

We fitted the XIS spectra in each region of the Hydra A cluster using a 
single-temperature 
APEC model for the ICM with the Galactic absorption, $N_{\rm H}$\@.
The spectra of each annular region of the
two fields were fitted simultaneously by minimizing the total $\chi^2$ value. 
Here, the common background model was included for all 
regions, where  the surface brightness of the background
components were restricted within the statistical errors derived for the outermost regions.
The temperature and normalization of the ICM component  were free parameters. 
The abundances in each annular region in the two fields
were assumed to have the same value.
$N_{\rm H}$ was fixed to the Galactic value of
$4.7\times 10^{20}~{\rm  cm^{-2}}$ in the direction of the Hydra~A cluster. 

The resultant parameters are summarized in Table~\ref{tab:results}, 
and the best-fit spectra are shown in Figure~\ref{fig:spec}. 
The spectra are well represented with the single-temperature model for the ICM
and the background.

\begin{table*}[th]
\caption{Parameters resulting from fitting ICM with the single-temperature APEC model  in the energy ranges of 0.7--7.0 keV.}
\label{tab:results}
\small
\tabcolsep 5pt
\begin{tabular}{cccccccc}
\hline \hline
 \multicolumn{2}{c}{Region} & \multicolumn{3}{c}{filament} & \multicolumn{2}{c}{void}\\
\hline
(arcmin) & (kpc) & $kT$ (keV) & abundance (solar) & normalization\footnotemark[$\ast$]
& $kT$ (keV)  & normalization\footnotemark[$\ast$] & $\chi^{2}$/d.o.f.\\
\hline
$2'-4'$ & $133-266$& $3.81_{-0.34}^{+0.35}$ & $0.33\pm0.10$ & $0.45_{-0.02}^{+0.02}$ &  $3.18_{-0.14}^{+0.15}$ & $0.36_{-0.02}^{+0.02}$ & $1.22(263/216)$ \\
$4'-6'$ & $266-400$ & $3.31_{-0.23}^{+0.32}$ & $0.43\pm0.43$ & $0.10_{-0.01}^{+0.01}$ & $3.09_{-0.24}^{+0.27}$ & $0.090_{-0.01}^{+0.01}$ & $1.09(146/134)$ \\
$6'-8'$ &$400-533$ & $3.11_{-0.23}^{+0.28}$ & $0.28\pm0.15$ & $0.056_{-0.005}^{+0.004}$ & $3.07_{-0.31}^{+0.38}$ &  $0.035_{-0.003}^{+0.003}$ & $1.18(177/150)$ \\
$8'-10'$ &$533-666$ & $2.60_{-0.27}^{+0.43}$ & $0.18\pm0.13$ & $0.025_{-0.003}^{+0.003}$ & $3.16_{-0.81}^{+0.86}$ & $0.015_{-0.002}^{+0.003}$ & $1.18(129/109)$ \\ 
$10'-12'$ & $666-799$& $2.54_{-0.36}^{+0.67}$ & $0.38\pm0.38$ & $0.013_{-0.002}^{+0.003}$ & $3.46_{-1.60}^{+1.60}$ &  $0.0058_{-0.0012}^{+0.0016}$ & $1.07(99/93)$ \\
$12'-15'$ & $799-997$& $2.36_{-0.44}^{+0.33}$ & $0.30\pm0.19$ & $0.010_{-0.001}^{+0.001}$  & $1.79_{-0.31}^{+0.33}$  & $0.0048_{-0.0012}^{+0.0012}$ & $1.13(170/150)$ \\
$15'-23'$ & $997-1529$& $1.44_{-0.33}^{+0.47}$ & $0.10\pm0.08$ & $0.0036_{-0.0011}^{+0.0012}$ &$1.39_{-0.42}^{+0.65}$  & $0.0025_{-0.0008}^{+0.0011}$ & $1.19(263/221)$ \\
\hline
\multicolumn{8}{@{}l@{}}{\hbox to 0pt{\parbox{170mm}{\footnotesize
\footnotemark[$\ast$] Normalization of the $APEC$, component divided by the solid angle, $\Omega^{\emissiontype{U}}$, 
assumed in the uniform-sky ARF calculation (\timeform{20'} radius), 
$Norm = \int n_{e}n_{H}dV/(4\pi(1+z)^{2}D_{A}^{2})/\Omega^{\emissiontype{U}}\times 10^{-14}$ cm$^{-5}$~400$\pi$~arcmin$^{-2}$, 
where $D_{A}$ is the angular distance to the source.
 }\hss}}
\end{tabular}
\end{table*}

\section{Results}

\subsection{Temperature profile}
\label{subsec:kT}

\begin{figure}[t]
\FigureFile(80mm, 60mm){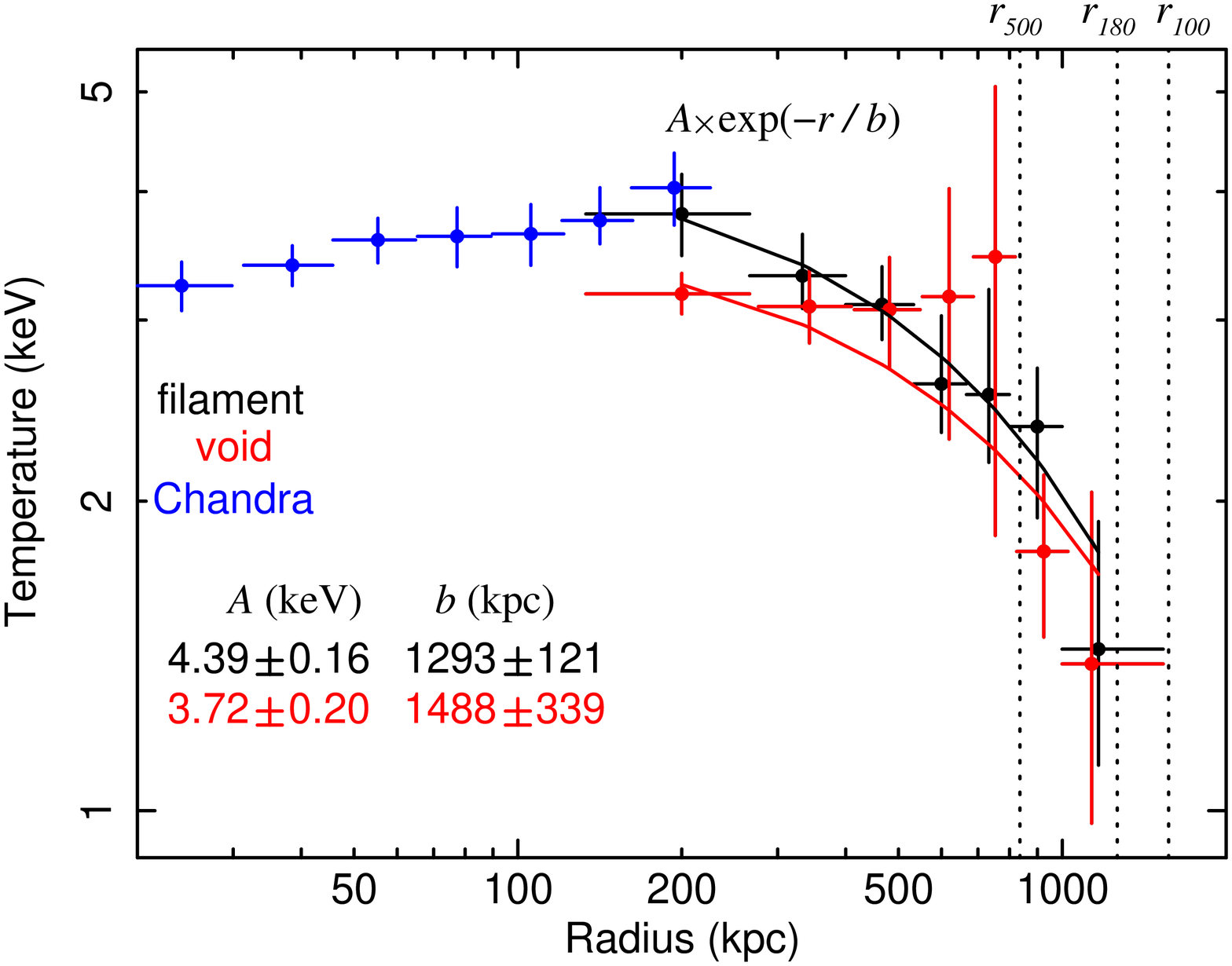}
\caption{
Radial temperature profiles of the filament (black) and void (red) 
directions. 
Solid lines show the best fit function of
$A \times \exp(-b \times r)$. Chandra results \citep{david01} 
are plotted in blue. 
Vertical dotted lines show $r_{500}, r_{180}$, and $ r_{100}$, 
derived from the fitting of the hydrostatic mass with 
the NFW universal mass profile \citep{navarro1997}, as described in subsection \ref{subsec:mass}.
}
\label{fig:kT}
\end{figure}

Figure \ref{fig:kT} shows the radial temperature profiles of the two
directions derived from the spectral fits.
At a given radius, the temperatures for each direction
are consistent within  statistical errors, except around $\sim$200 kpc 
 from the center. 
The region around 200 kpc corresponds to the shock front detected
by Chandra and XMM in the filament direction
\citep{simonescu09b, gitti11}.
From 300 kpc to 1200 kpc, the temperatures in each direction 
decrease with radius in a similar manner down to $\sim$ 1.4 keV.

From ROSAT observations,
\citet{ikebe97} found that the spectrum at the central region within 
2$'$ exhibits  the multiphase feature.  Although the 
central 2$'$ region was excluded in our spectral analysis, we examined 
the spectral fits with the two-temperature model in  regions with $r~>~2'$.
The  reduced-$\chi^{2}$ for  the two-temperature model was not 
significantly improved in comparison  with the single-temperature model.

To determine the hydrostatic mass of Hydra A, we fitted 
the temperature profile of each direction with the 
exponential formula, $A \times \exp(-b\times r)$, where $A$ and $b$ 
are free parameters with units of keV and kpc$^{-1}$, respectively,
and $r$ is the distance from the X-ray peak of the cluster center.
The resulting fit parameters and best-fit functions
 are shown in Figure \ref{fig:kT}. 
In the filament direction,
the normalization factor $A$ of the exponential model
 was 18\% higher than that in the void direction, which
reflects a temperature difference of $\sim$ 200 kpc.
The parameter $b$ for both directions is consistent within  
statistical error.

\subsection{Deprojected electron density profile}
\label{subsec:ne}

\begin{figure}[t]
\FigureFile(80mm, 60mm){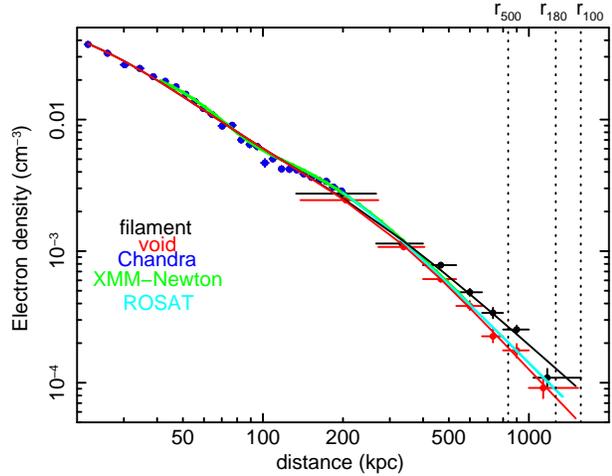}
\caption{
Deprojected electron density profile of the filament (black) 
and void (red) directions. Black and red solid lines show the best fit 
with the double-$\beta$ model.
Green solid line shows the electron density profile derived 
with XMM data, the light blue solid line shows that 
 derived from the ROSAT surface-brightness
profile \citep{ikebe97} scaled with the XMM profile, and 
the blue filled circles show that derived from 
Chandra results \citep{david01}.
}
\label{fig:ne}
\end{figure}

\begin{table*}[th]
\caption{Parameters resulting from fitting  the deprojected density profile 
derived from XMM and Suzaku
with the double-$\beta$ model\footnotemark[$*$].}
\label{tab:nefit}
\begin{center}
\begin{tabular}{ccccccc}
\hline \hline
 & \multicolumn{3}{c}{narrower component} & \multicolumn{3}{c}{wider component}\\
\hline
Field & $n_{0,1}$ (10$^{-2}$~cm$^{-3}$) & $r_{c,1}$ (kpc) & $\beta_{1}$ & $n_{0,2}$ (10$^{-3}$~cm$^{-3}$) & $r_{c,2}$ (kpc) & $\beta_{2}$\\
\hline
void & $6.73\pm0.17$ & $21.3\pm4.7$ & $0.59\pm0.01$ & $2.16\pm0.14$ & $278\pm264$ & $0.95\pm0.07$\\
filament & $6.61\pm0.19$ & $22.3\pm5.4$ & $0.62\pm0.01$ & $2.73\pm0.15$ & $193\pm114$ & $0.60\pm0.03$\\
\hline
\multicolumn{7}{@{}l@{}}{\hbox to 0pt{\parbox{170mm}{\footnotesize
\footnotemark[$*$] double-$\beta$ model is given by the form 
$n(r) = n_{0,1}(1+(r/r_{c,1})^2)^{-3\beta_{1}/2}+n_{0,2}(1+(r/r_{c,2})^2)^{-3\beta_{2}/2}$.\\
The errors of $n_{0,1}$ and $\beta_{1}$ are calculated by fixing the $r_{c,1}$ and wider component,
and vice versa.
\par\noindent
 }\hss}}
\end{tabular}
\end{center}

\caption{Resultant fit parameters of the deprojected density profile 
derived from XMM and Suzaku with the single-$\beta$ model\footnotemark[$*$]
beyond 100 kpc.}
\label{tab:nesinglefit}
\begin{center}
\begin{tabular}{cccc}
\hline 
\hline
Field & $n_{0}$ (10$^{-3}$~cm$^{-3}$) & $r_{c}$ (kpc) & $\beta$ \\
\hline
void & $7.38\pm0.42$ & $160\pm13$ & $0.74\pm0.03$ \\
filament & $7.98\pm0.50$ & $131\pm10$ & $0.61\pm0.02$ \\
\hline
\multicolumn{4}{@{}l@{}}{\hbox to 0pt{\parbox{170mm}{\footnotesize
\footnotemark[$*$] single-$\beta$ model is given by the form 
$n(r) = n_{0}(1+(r/r_{c})^2)^{-3\beta/2}.$\\
\par\noindent
 }\hss}}
\end{tabular}
\end{center}
\end{table*}

We calculated three-dimensional electron density profiles from the normalization of
the ICM component derived with Suzaku, 
considering the geometrical volume that contributes to each
two-dimensional region.
Here,  we assumed spherical symmetry within each field.
The electron density profile was also obtained using XMM data.
To avoid uncertainties in the background,
we fitted the annular spectra of XMM MOS with the APEC model in an energy range of 
1.6--5.0 keV.
Here, the temperature and Fe abundance were left free within the error bars
derived by \citet{matsushita11}.
We fitted the derived emissivity profiles with a double-$\beta$-model,
and calculated the three-dimensional radial profile of the electron-density.

Figure \ref{fig:ne} shows the electron density profile derived from XMM and Suzaku (this work), 
Chandra \citep{david01} and the ROSAT  surface-brightness profile
scaled with the XMM density profile
 at 200 kpc \citep{ikebe97}.
The deprojected electron density profile derived from XMM and Suzaku 
observations agree well with those from ROSAT within 500 kpc and from Chandra within 200 kpc.

At a given distance from the X-ray peak,
electron density beyond 400 kpc in the filament direction 
was systematically higher than that in the void direction,
reflecting the difference in  normalizations
(Table \ref{tab:results} and  Figure \ref{fig:spec}).
At 900 kpc, for example, the deprojected electron densities in the filament and void 
directions were $(2.54\pm0.18)\times10^{-4}$~cm$^{-3}$ and 
$(1.76\pm0.21)\times10^{-4}$~cm$^{-3}$, respectively.
The electron density dropped to 
$\sim$10$^{-4}$~cm$^{-3}$ around 1200 kpc.

We fitted the density profiles from XMM and Suzaku results with 
the double-$\beta$ model from 40 to 1200 kpc.
The  results are shown in Figure \ref{fig:ne}, and the resulting
parameters are given in Table \ref{tab:nefit}. 
From 400 kpc to 1530 kpc,
the radial profiles of the electron density are also well represented 
by a power--law profile with slopes of 
$\Gamma=1.60\pm0.05$ and $1.72\pm0.07$ in the filament and void 
directions, respectively. 
The slope of
the electron density in the void direction was  slightly
steeper than that in the filament direction.
Fitting the electron density profiles beyond 100 kpc with the
single $\beta$-model resulted in $\beta=$ 0.61 $\pm$ 0.02 and $\beta=$ 0.74 $\pm$ 0.03 
for the filament and void directions, respectively.
These values are close to the value of 0.68
 derived from the ROSAT surface-brightness profile  \citep{ikebe97}.

\subsection{Entropy profiles}
\label{subsec:ent}

\begin{figure}[t]
\FigureFile(80mm, 60mm){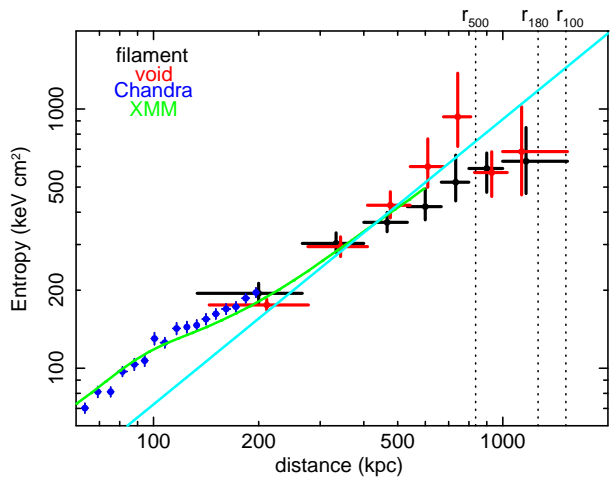}
\caption{
Entropy profiles in the filament (black) and void (red) directions.
Crosses and diamonds show entropy calculated from the 
resulting temperature and electron density, respectively, as shown in Figures 4 and 5.  
Chandra results \citep{david01} are represented by blue crosses.
The green line shows the entropy profile calculated by 
combining the electron density from XMM and the temperature 
from Suzaku.
The light blue line extrapolates the Suzaku and 
XMM data (green line) with a power-law formula fit for the data beyond 200 kpc, 
which has a fixed index of 1.1. 
}
\label{fig:ent}
\end{figure}

Entropy, which is a  useful tool for investigating the thermodynamic 
history of hot gas in groups and clusters, is defined as 
\begin{equation}
S = \frac{kT}{n_{e}^{2/3}},
\end{equation}
where $T$ and $n_{e}$ are the temperature and deprojected electron density, respectively.
In Figure \ref{fig:ent}, we show the entropy profiles of the Hydra A
cluster calculated from the
derived temperature and electron density profiles shown in 
Figures \ref{fig:kT} and \ref{fig:ne}.
Entropy in the two directions increased from the center to 
800 kpc, or $r_{500}$, which was derived from the Navarro, Frenk ans White(NFW) model fit
to the hydrostatic mass described in section \ref{subsec:mass}. 
Between 500 kpc and 800 kpc, the entropy in the void direction 
was systematically higher than  in the filament direction.
At  the two outermost radial bins (beyond $r_{500}$), the entropies 
of the two directions agree well.
In addition, we calculated the entropy profile by combining the 
parameters resulting from the fit of the electron density profile from XMM and 
the temperature profile from Suzaku as indicated by the green line
in Figure \ref{fig:ent}. 
The entropy profiles using Suzaku, XMM, and
Chandra data are consistent with each other (figure \ref{fig:ent}).

The derived entropy profiles were compared 
with a power-law model with a fixed index of 1.1, which 
was expected from simulations of the accretion shock heating model
(\cite{tozzi01}; \cite{ponman03}).
From 200 kpc to 500 kpc, the derived entropy profiles agreed well
with the  $r^{1.1}$ relationship.
In contrast,
beyond 800 kpc, or $\sim r_{500}$, the entropy profiles in the two directions were
systematically lower than the predicted by the $r^{1.1}$ relationship.
Inside 200 kpc, the entropy exceeded the model prediction,
which attribute to the presence of the cool core boundary 
\citep{peres1998} and the shock front around 200 kpc.

\subsection{Hydrostatic mass}
\label{subsec:mass}
\begin{figure*}[t]
\FigureFile(80mm, 60mm){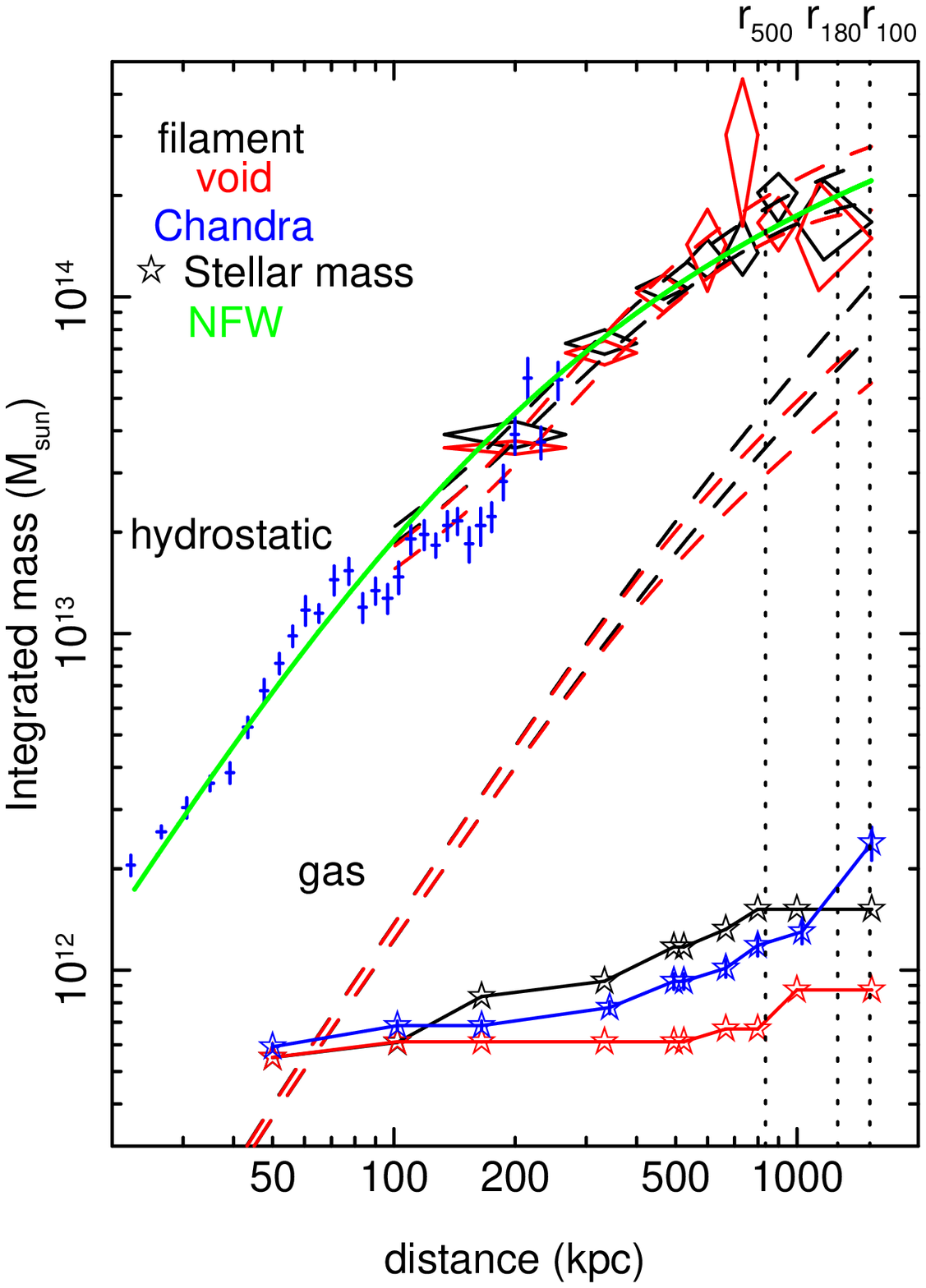}
\FigureFile(80mm, 60mm){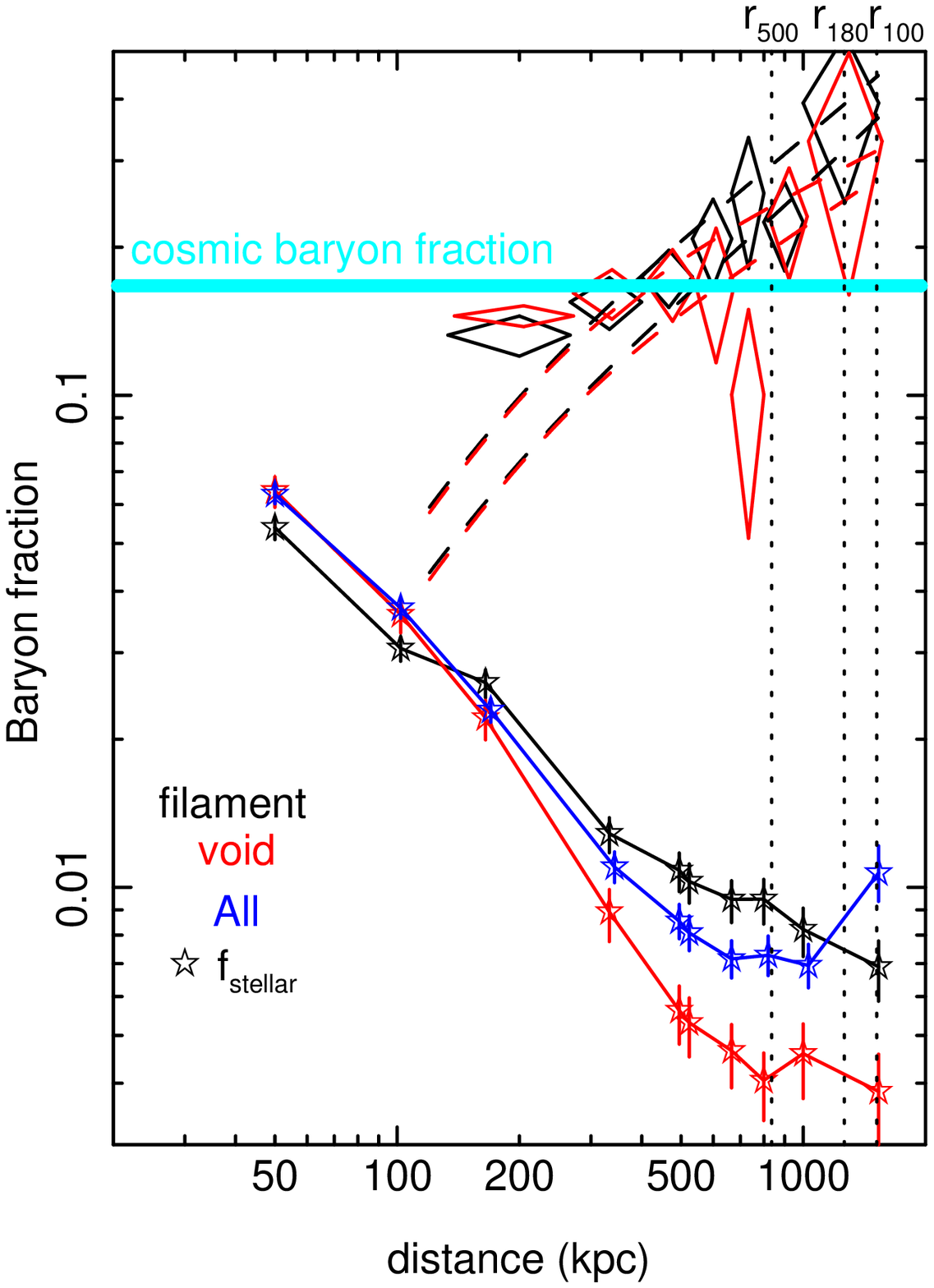}
\caption{
left:Integrated hydrostatic and gas-mass profiles. 
All  the black and red colors show the filament and void directions, respectively.
The masses were 
calculated in the whole azimuthal angle under the assumption of 
spherical symmetry for each direction. 
The hydrostatic mass was calculated from 
the temperatures of shells $(i-1)$, $i$, and $(i+1)$ (diamonds)
and the exponential relationship (dashed lines).
All dashed lines show 90\% upper and lower limits.
The green line shows the best-fit NFW model.
The integrated stellar mass profiles in the filament and void directions
using 2MASS data are represented by  stars.
Blue stars are the integrated stellar mass profile
in the whole azimuthal angle corrected for the contribution of the
fainter galaxies using LF (solid).
right:Cumulative profiles of gas fraction,
 $f_{\rm gas}(<r)$ assuming 
hydrostatic equilibrium in  both directions. 
The diamonds show $f_{\rm gas}(<r)$ of the integrated hydrostatic and gas mass
of the corresponding directions.
Dashed lines represent $f_{\rm gas}(<r)$ applying
the result of the NFW model fit.
The value of the cosmic baryon fraction (light blue) from 
WMAP \citep{komatsu11} is plotted.
The stars represent the stellar fraction, $f_{\rm star}(<r)$,
using the NFW model fit.
The meanings of the colors and line styles are
the same as in the left panel.
}
\label{fig:fbaryon}
\end{figure*}

We estimated the hydrostatic or gravitational mass assuming 
the hydrostatic equilibrium given by,

\begin{equation}
\label{eq:hydro}
M_{\rm H.E.}(< r) = -\frac{kT(r)r}{\mu m_{p}G}\left( 
\frac{d \ln \rho (r)}{d \ln r} + \frac{d {\ln} kT(r)}{d \ln r}\right),
\end{equation}
where $M_{\rm H.E.}(< r)$ is
the hydrostatic mass within the three-dimensional radius $r$,
 $G$ is the gravitational constant, $k$ is the Boltzmann constant, 
$\rho$ is the gas density, $kT$ is the temperature, and $\mu m_{p}$ is 
the mean particle mass of the gas (the mean molecular weight is $\mu=0.62$).
Here, we assumed spherically symmetric  mass distribution.

On the basis of the temperature and electron density profiles,
we calculated the hydrostatic mass of the Hydra A cluster beyond 100 kpc.
The temperature gradient of the $i$th shell was calculated
from a power-law fit of temperatures of shells $(i-1)$, $i$, and $(i+1)$.
We employed 
the electron density profile from the single-$\beta$ model 
fit shown in table \ref{tab:nesinglefit}.
The  derived hydrostatic mass is plotted 
in Figure \ref{fig:fbaryon} as diamonds with the integrated gas mass profiles and
stellar mass profiles derived in subsection 4.6.
The derived mass was 
calculated for the whole azimuthal angle under the assumption of  spherical symmetry using
the parameters derived for each direction: thus,
the mass in each direction was four times larger than that in 
the corresponding direction.
Although the electron density  in the filament direction at a given radius
was higher than that in the void direction, 
the difference in the slopes of the electron density was small.
As a result, 
the hydrostatic masses derived for the two directions agree well.
In addition, we calculated the 90\% range of hydrostatic mass using
the exponential relationship of the temperature profiles and fits
of the electron-density profiles with 
the double-$\beta$ model, 
which are plotted as dashed lines in Figure \ref{fig:fbaryon}.


We fitted the derived hydrostatic mass profiles
with the following NFW  formula, which shows the equilibrium 
density profile of the dark matter halo \citep{navarro1996, navarro1997};
\begin{equation}
M_{\rm NFW}(< r) = 4\pi\delta_{c}\rho_{c}r_{s}^{3}\left[\ln(1+x) - 
\frac{x}{1+x}\right], x\equiv r/r_{s},
\end{equation}
where $M_{\rm NFW}(< r)$ is the mass within a radius $r$,
 $\rho_{c}$ is the critical baryon density of the Universe, 
$r_{s}$ is the scaled radius, and $\delta_{c}$ is the characteristic 
density, that can be expressed in terms of the concentration parameter
($c = r_{200}/r_{s}$) as
\begin{equation}
\delta_{c} = \frac{200}{3}\frac{c^{3}}{\ln(1+c) - c/1+c}.
\end{equation}
Since the  hydrostatic masses  derived for  the two directions
agree well, we fitted the two hydrostatic mass
profiles with Suzaku and that with Chandra simultaneously,
excluding the shock region from 100 kpc to 200 kpc.
The parameters derived from the fit are shown in Table \ref{tab:nfw}.
The quantity $r_{100}$ is equal to $r_{vir}$ in Hydra A cluster, since
for our  fiducial cosmological model, $\Delta_{vir} = 100$ for halos with a 
redshift $z = 0.0539$ \citep{nakamura1997}. We estimated $r_{\Delta}$ for various 
overdensities: $\Delta =500,\ 200,\ 180$ and 100,
and $r_{\Delta}$ derived from fits with the NFW model are 
$811\pm10$ kpc, $1189\pm96$ kpc, $1243\pm16$ kpc, and $1577\pm22$ kpc, respectively.
The $r_{100}$ is close to the outermost ring radius.
Possible deviations from hydrostatic equilibrium have been previously discussed
to explain the entropy flattening beyond $r_{500}$ 
\citep{bautz09, george09, kawaharada10}: therefore,
we fitted the hydrostatic mass out to $r_{500}$.
We were able to obtain almost the same results as shown in Table \ref{tab:nfw}.
Thus, the NFW model represents the hydrostatic mass out to 
$r_{100}$ within error bars as shown in Figure \ref{fig:fbaryon}.

\begin{figure}[t]
\FigureFile(80mm, 60mm){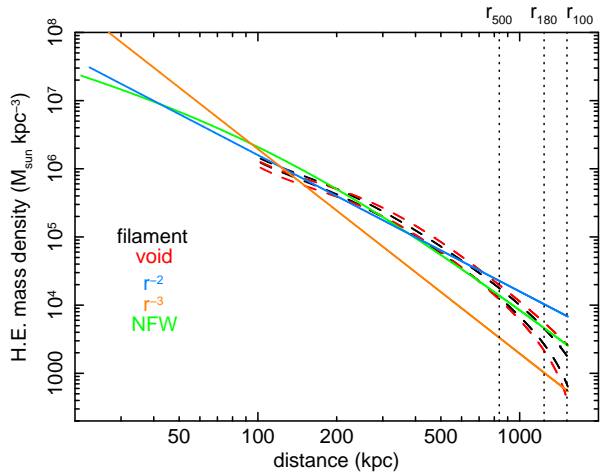}
\caption{
H.E. mass density profiles for 
the filament (black) and void (red) directions.
Density was calculated from the integrated H.E. mass profile.
The $r^{-2}$ and $r^{-3}$ profiles are plotted with solid blue and orange
lines, respectively.
}
\label{fig:mass}
\end{figure}

Moreover, we calculated the total mass density profile from the total 
integrated hydrostatic mass profile in each direction.
The density profile was also calculated from the NFW model obtained above. 
The calculated density profiles were compared with $r^{-2}$ and $r^{-3}$ profiles
(Figure \ref{fig:mass}).
The NFW density profile varied from $r^{-1}$ to $r^{-3}$ with increasing radius.
Within $r_{500}$, the densities are well represented by the $r^{-2}$ profile,
which is shallower than the asymptotic matter density slope $\rho \propto r^{-3}$.
Beyond $r_{500}$, the $r^{-3}$ profile is preferable than the $r^{-2}$ profile.

\begin{table*}[th]
\caption{
Cumulative values of the hydrostatic mass ($M_{H.E.}$), gas mass ($M_{gas}$), and gas fraction 
($f_{gas}$) of the Hydra Hydra A cluster. }
\label{tab:mass}
\small
\tabcolsep 4pt
\begin{center}
\begin{tabular}{ccccccc} 
\hline
Region & \multicolumn{3}{c}{filament} & \multicolumn{3}{c}{void} \\
\hline
$r$ & $M_{gas}$ & $M_{H.E.}$ & $f_{gas}$ & $M_{gas}$ & $M_{H.E.}$ & $f_{gas}$ \\
(kpc) & $(10^{14}~M_{\solar})$ & $(10^{14}~M_{\solar})$ & & $(10^{14}~M_{\solar})$ & $(10^{14}~M_{\solar})$ \\
\hline
$133-266$ & $0.052\pm0.001$ & $0.39\pm0.04$ & $0.13\pm0.01$ & $0.052\pm0.001$ & $0.36\pm0.02$ & $0.15\pm0.01$ \\
$266-400$ & $0.11\pm0.01$ & $0.72\pm0.07$ & $0.15\pm0.02$ & $0.11\pm0.01$ & $0.68\pm0.06$ & $0.16\pm0.02$ \\
$400-533$ & $0.19\pm0.02$ & $1.06\pm0.09$ & $0.17\pm0.02$ & $0.17\pm0.02$ & $1.03\pm0.13$ & $0.17\pm0.03$ \\
$533-666$ & $0.26\pm0.03$ & $1.27\pm0.21$ & $0.21\pm0.04$ & $0.24\pm0.03$ & $1.43\pm0.39$ & $0.17\pm0.05$ \\
$666-799$ & $0.35\pm0.05$ & $1.35\pm0.35$ & $0.26\pm0.08$ & $0.30\pm0.04$ & $3.03\pm1.41$ & $0.10\pm0.05$ \\
$799-997$ & $0.46\pm0.07$ & $2.04\pm0.37$ & $0.22\pm0.05$ & $0.38\pm0.06$ & $1.66\pm0.31$ & $0.23\pm0.06$ \\
$997-1529$ & $0.71\pm0.12$ & $1.82\pm0.60$ & $0.39\pm0.14$ & $0.55\pm0.11$ & $1.67\pm0.78$ & $0.33\pm0.17$ \\
\hline
\end{tabular}
\end{center}
\end{table*}

\begin{table*}[th]
\caption{
Resulting radii $r_{c}$ and $r_{200}$, and
concentration parameter $c_{200}$ derived from the NFW model.
}
\label{tab:nfw}
\small
\tabcolsep 4pt
\begin{center}
\begin{tabular}{cccc} 
\hline
 & $r_{s}$ (kpc) & $r_{200}$ (kpc)& $c$ \\
\hline
fitted up to $r_{500}$ & $126\pm8$  & $1189\pm96$ & $9.44\pm0.48$ \\
fitted up to $r_{100}$ & $123\pm8$  & $1183\pm98$ & $9.62\pm0.49$ \\
\hline
\end{tabular}
\end{center}
\end{table*}

\subsection{Gas fraction}
\label{subsec:fraction}

We derived the cumulative gas mass fraction as,
\begin{equation}
f_{\rm gas}(< r) = \frac{M_{\rm gas}(< r)}{M_{\rm H.E.}(< r)},
\end{equation}
where $M_{\rm gas}(< r)$ and $M_{\rm H.E}(< r)$ are the gas mass and hydrostatic mass,
respectively, within a sphere of radius $r$.
As shown in Figure \ref{fig:fbaryon} and Table \ref{tab:mass},
the gas fraction increased in the outskirts
and reached the cosmic mean baryon fraction derived from
seven-year data of Wilkinson microwave anisotropy probe
 (WMAP7; \cite{komatsu11}) at $\sim 0.5~r_{500}$.
At $r_{100}$, $f_{\rm gas}(<r)$ in the filament direction  
exceeded the result of WMAP7,
while the lower limit of $f_{\rm gas}(<r)$  in the void direction reached
 the WMAP 7 fraction.
Adopting the fit of the NFW model to the hydrostatic mass resulted in
 $f_{\rm gas}(<r)$ in the void and the filament directions at $r_{100}$,
exceeding the cosmic baryon fraction by a factor of 2 and 3, respectively.
Furthermore,
$f_{\rm gas}(<r)$ in the void direction  at $r_{180}$ was close to
that derived for the north-west direction of the Perseus cluster at $r_{200}$
\citep{simionescu11}:
$r_{180}$ was derived from the NFW model fit of the hydrostatic mass.
The radial profile of $f_{\rm gas}(<r)$ of the Abell 1246 cluster with $kT=6.0$ keV
exhibited similar behavior as that of the our result of the
Hydra A cluster \citep{sato12}.

\subsection{K-band Luminosity of galaxies and stellar fraction}
\label{subsec:imlr}

\begin{figure*}[t]
\FigureFile(80mm,60mm){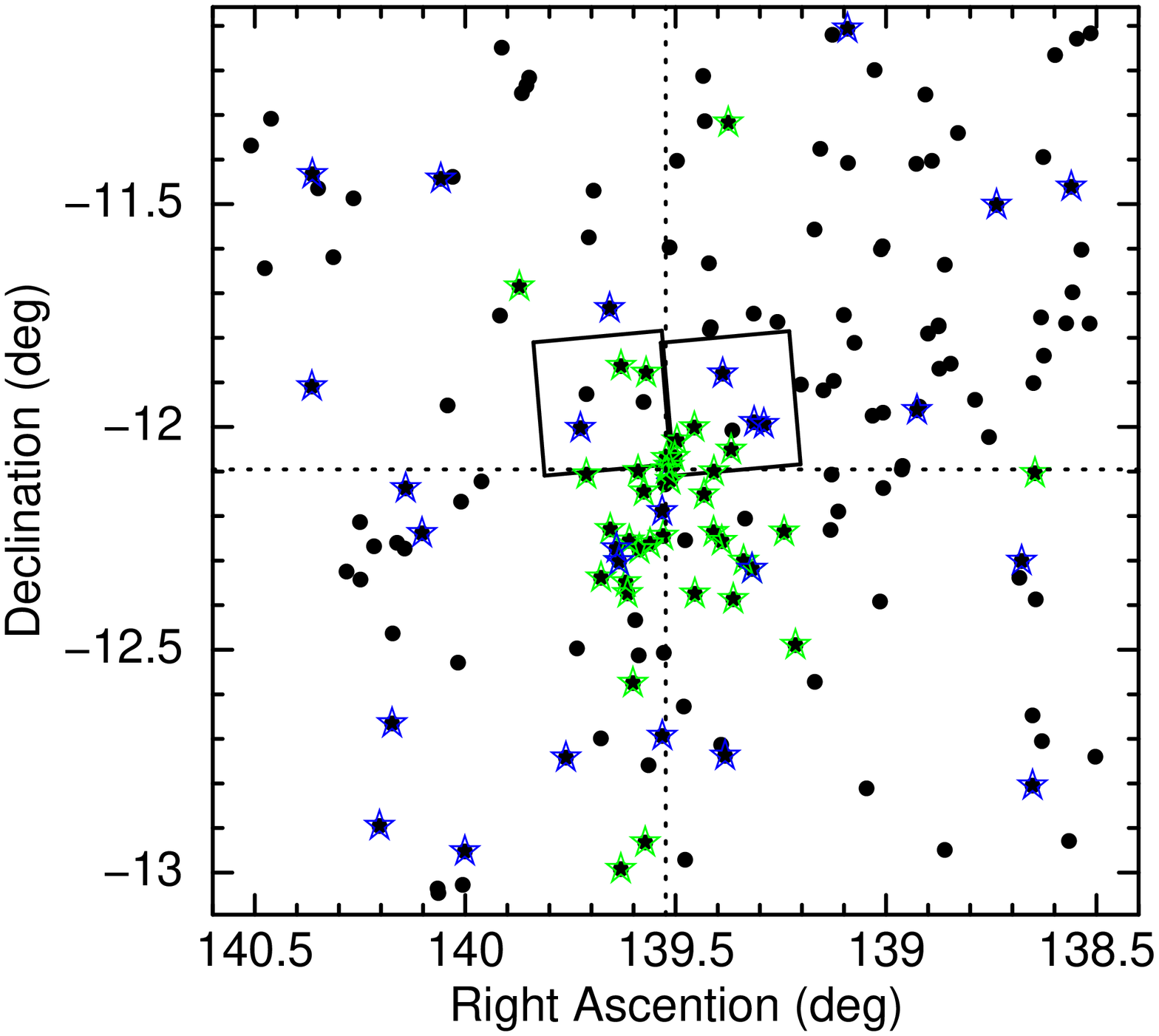}
\FigureFile(72mm,54mm){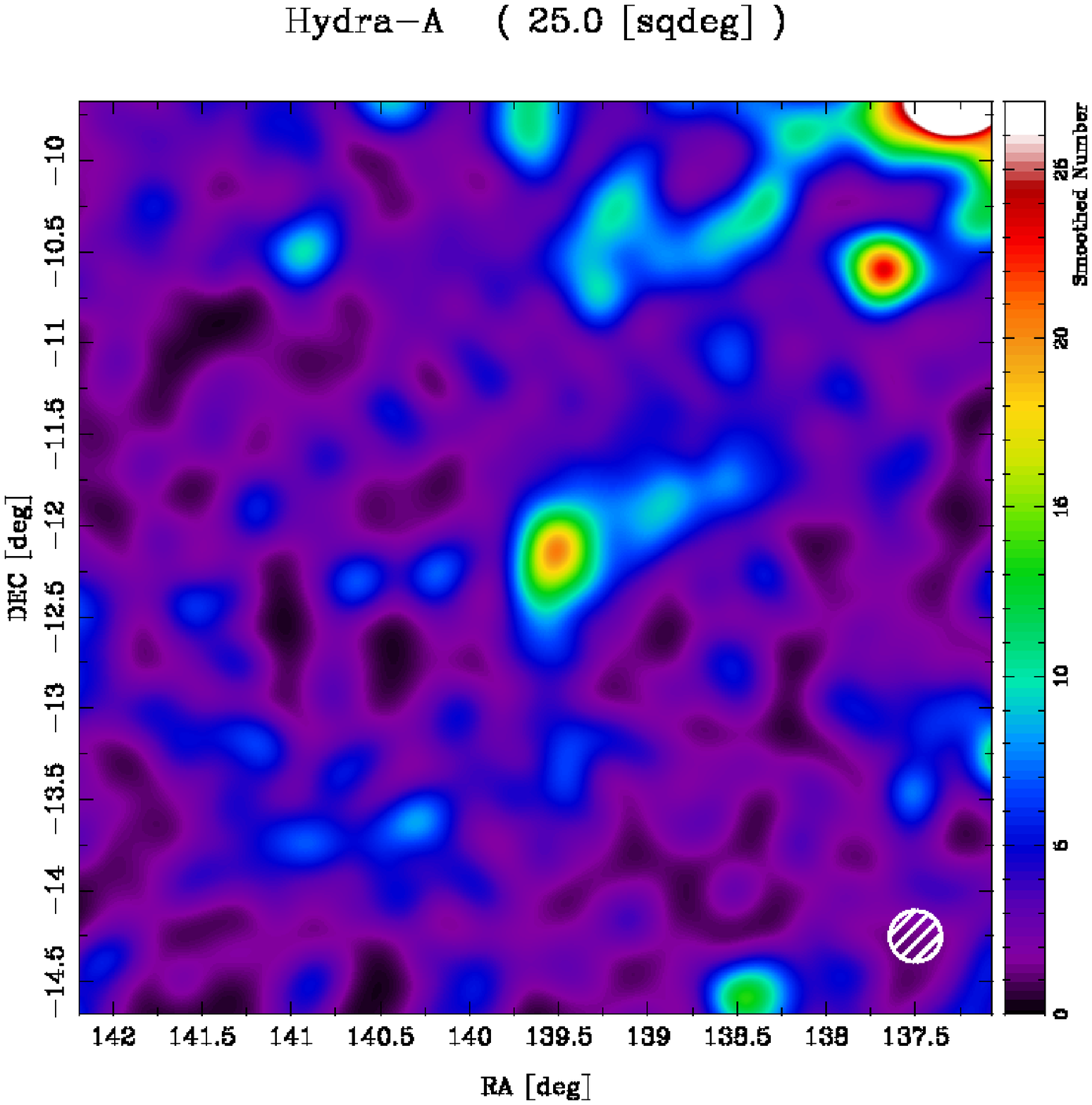}
\caption{
Two-dimensional density distributions of galaxies around Hydra A,
retrieved from 2MASS in the $K$-band. 
left: 
Stars represent galaxies with available redshift
in the NASA/IPAC Extragalactic Database (NED).
Green and blue stars indicate members and non-members, respectively,
 of the Hydra A cluster based on their redshift.
The boxes represent the Suzaku fields of view (FOVs) of XIS pointings. 
The directions of the filament (north-west) and the void (north-east)
are separated as dashed lines.
Right: The box size is \timeform{5D}$\times$\timeform{5D}.
Abell 754 cluster, with coordinates (RA, Dec) = (137.2087, -9.6366) in degrees, 
is shown in top right.
A Gaussian smoothing scale is represented by the white hatched circle 
at bottom right.
}
\label{fig:extended source}
\end{figure*}

\begin{figure*}
\FigureFile(80mm,60mm){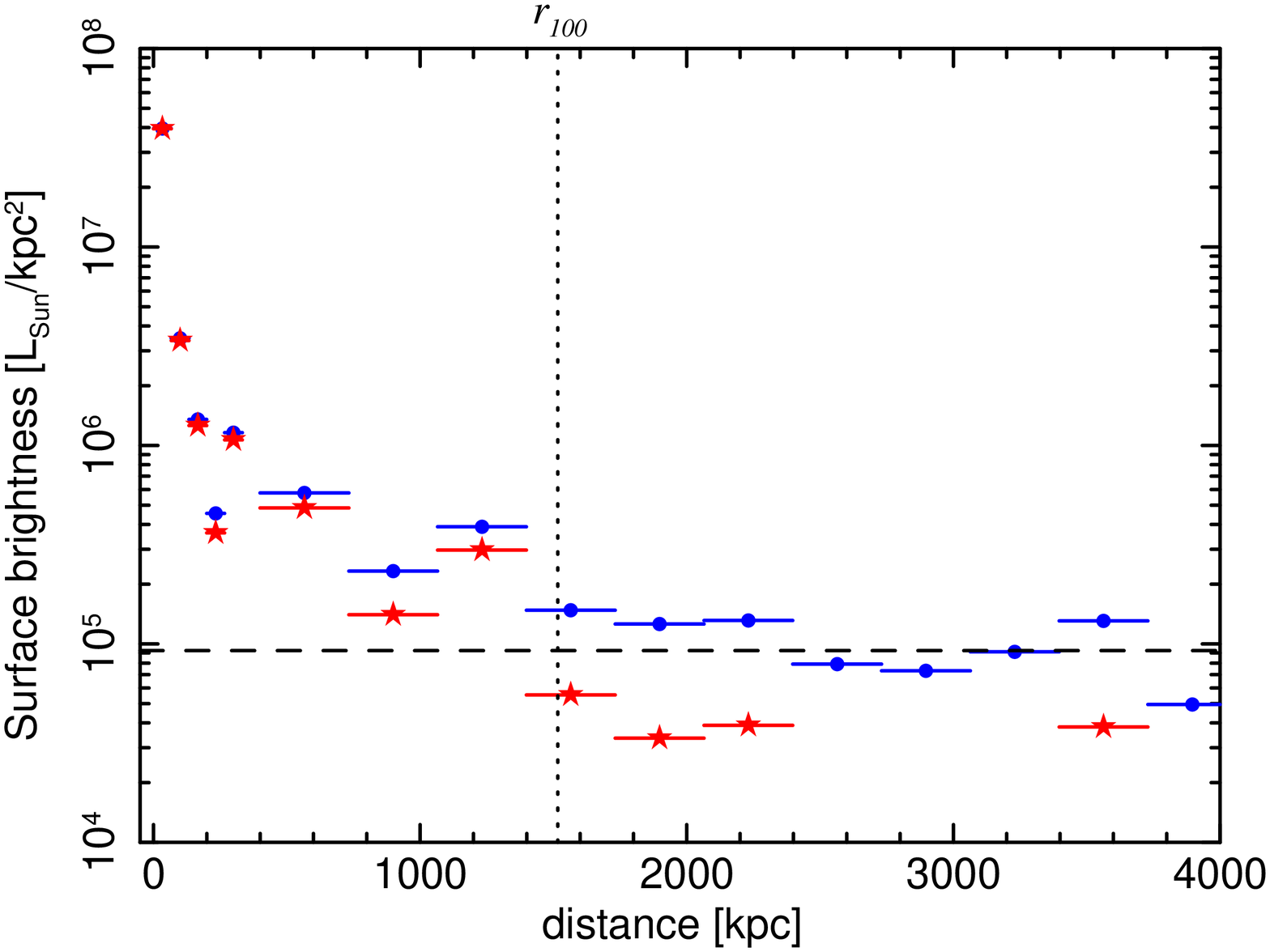}
\FigureFile(80mm,60mm){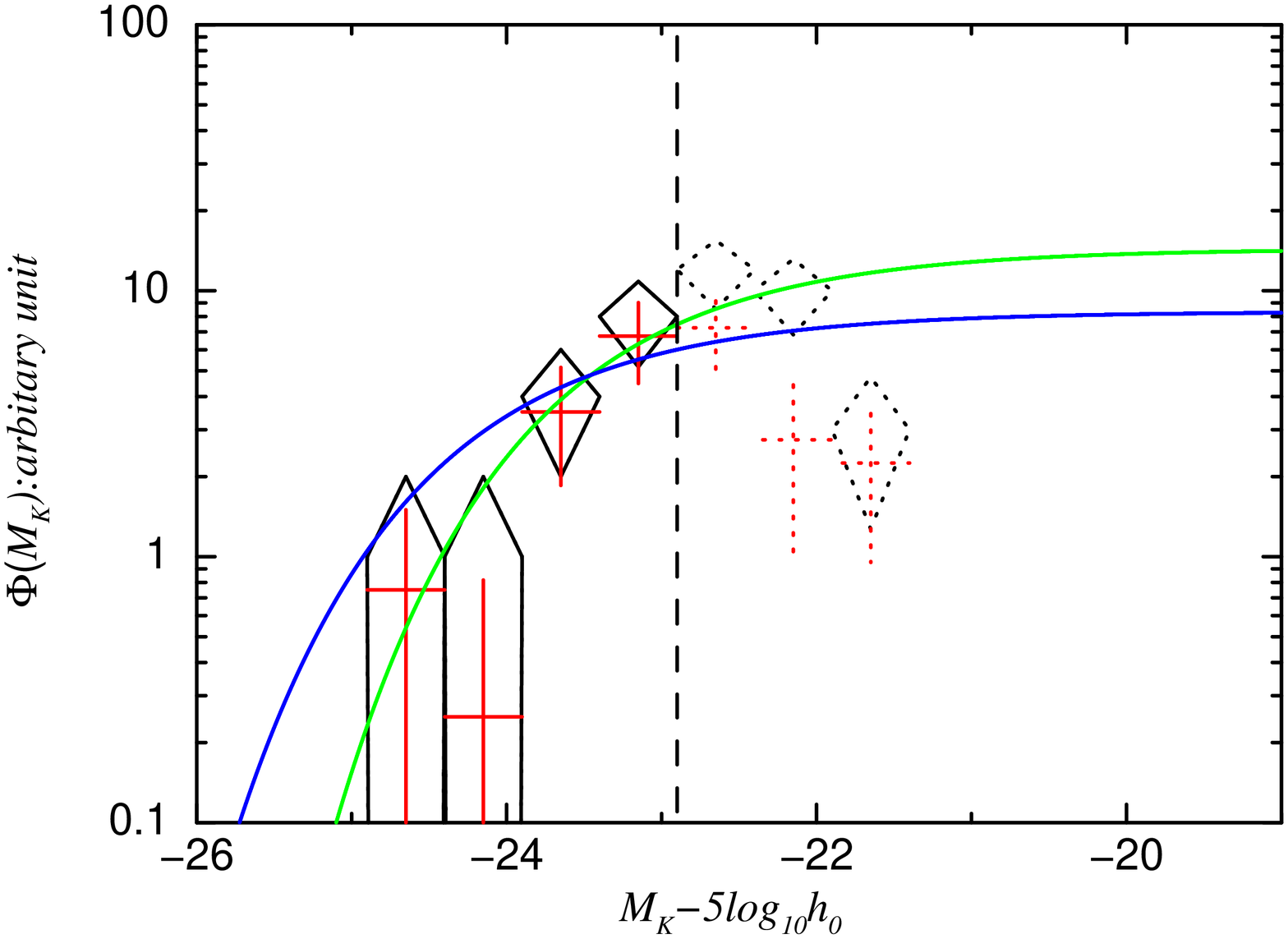}
\caption{left:Surface-brightness profile of Hydra A
observed with 2MASS photometric data in the $K$-band. 
Blue points denote the detected galaxies of Hydra A
and red stars denote the background-subtracted data.
The dashed line shows the subtracted background level
and the dotted line shows $r_{100}$.
right: The LF of Hydra A observed with 2MASS photometric data in the $K$-band 
within $r_{100}$ in absolute magnitude form, including background galaxies (black diamonds) and 
that the subtracted background contribution (red crosses).
The vertical dashed line represents the complete limit of 2MASS.
Blue and green solid lines show the best-fit Schechter functions
when $L_*$ was fixed to the result  from the stacked LF of
 galaxies in clusters by \citet{lin03} and allowed to vary, respectively.
}
\label{fig:sfandlf}
\end{figure*}

The ratio of gas mass to stellar luminosity is a key parameter
for studying the star formation efficiency in clusters of galaxies.
Since almost all  metals in the ICM are synthesized in galaxies, the
metal-mass-to-light ratio provides  a useful measure for studying the
ICM chemical evolution.
To estimate the gas-mass and Fe-mass-to-light ratios,
we calculated the $K$-band luminosity of Hydra A cluster on the basis of 
the Two Micron All Sky Survey (2MASS) catalog,
whereby  all data is presented in a \timeform{2D}$\times$\timeform{2D} box: 
the coordinates of objects are shown in the left panel of Figure \ref{fig:extended source}.
The smoothed distribution of galaxies around Hydra A obtained from 2MASS
is shown in the right panel in Figure \ref{fig:extended source}.
The distribution of galaxies is elongated in the filament direction 
toward the Abell 754 cluster, which exhibited a redshift close to that of Hydra A.
In the void direction, the density of galaxies was significantly lower.
Toward the south, an additional filament-like distribution of galaxies appeared from 
the Hydra A cluster.
Hydra A itself was discovered to have an apparent magnitude of $m_{K}=11.07$, or
$L_{K}=5\times 10^{11}L_{K,\odot}$ 
 using a luminosity distance $D_{L}=240$ Mpc and a
foreground Galactic extinction of $A_{K}$ = 0.015 \citep{schlegel1998}.
Among galaxies 
in the NED data base\footnote{http://nedwww.ipac.caltech.edu/}
with available redshifts,
those with heliocentric velocities
differ from that of the Hydra A cluster by $>$2000 km/s were excluded from the analysis.
The $K$-band surface brightness profile centered on Hydra A
 is shown in figure \ref{fig:sfandlf}.
Beyond $\sim$2000 kpc from the cluster center, the $K$-band surface brightness flattened. 
Therefore,
we adopted the region between 2000 kpc and 4000 kpc as a background.

Figure \ref{fig:sfandlf} also shows the luminosity function (LF) of galaxies
in the Hydra A cluster within $r_{100}$.
Here, we used as background the LF of galaxies from 2000 kpc to 4000 kpc from the center of the
Hydra A cluster. 
The completeness limit of galaxies of 2MASS is $K_s$ = 13.5 in apparent magnitude
\footnote{See http://www.ipac.caltech.edu/2mass/releases/second/doc/explsup.html},
which corresponds to the $K$-band luminosity, $L_{K}$=$7\times 10^{10}L_{K,\odot}$ 
for the distance of the Hydra A cluster. 
The total $K$-band luminosity of galaxies above the 2MASS complete limit 
and that of all the detected galaxies including  those below the limit are
 1.5 $\times 10^{12}L_{K,\odot}$ and  2.5 $\times 10^{12}L_{K,\odot}$, respectively.
To calculate the total light of the cluster, we evaluate the contribution of fainter
galaxies below the 2MASS limit.
The LF of galaxies in clusters can be described with the Schechter function as,
\begin{equation}
\phi(L) dL = \phi_*\left(\frac{L}{L_*}\right)^\alpha\exp(-L/L_*)d\left(\frac{L}{L_*}\right)
\end{equation}
where the parameters $L_*$, $\phi_*$, and $\alpha$ are the characteristic luminosity, number
densities, and faint-end power-law index, respectively.
Upon adopting the $L_*$ of stacked cluster galaxies by \citet{lin03} and
integrating the Schechter function below the 2MASS limit,
 the total $K$-band luminosity becomes 2.2 $\times 10^{12}L_{K, \odot}$, which is
smaller than that of all the galaxies detected by 2MASS including those below the complete limit.
If we allow  $L_*$ to vary, the fit  for the LF of the Hydra A cluster improves,
as shown in Figure \ref{fig:sfandlf}, and the total K-band luminosity becomes
 2.8 $\times 10^{12}L_{K, \odot}$.
Therefore, the simple sum of $K$-band luminosities of the 
detected galaxies, including those below the 2MASS complete limit
is close to that integrated luminosity of the Schechter function.
Therefore, we adopted the total $K$-band luminosity of (2.5 $\pm$ 0.3) $\times 10^{12}L_{K, \odot}$.

We collected $K$-band luminosity of galaxies in the north-west and
north-east  sectors for the filament and void directions, respectively,
as well as  in the whole azimuthal angle and derived brightness profiles in the $K$-band.
For the whole azimuthal angle, we collected galaxies above the 2MASS complete limit 
and evaluated the contribution of fainter galaxies by integrating the Schechter function.
We obtained almost the same radial profile of $K$-band luminosity as
that from the  simple sum of luminosities of all the galaxies detected  with 2MASS.
Therefore, for the filament and void directions, we collected
all the galaxies detected by 2MASS in each direction
and derived radial profile of K-band luminosity.
After subtracting the background, we deprojected the brightness profile of
galaxies (under the  assumption of  spherical symmetry)
and derived three-dimensional radial profiles of $K$-band luminosity
in the filament, void, and  whole azimuthal angle.

Figure \ref{fig:fbaryon} shows the integrated $K$-band luminosity profiles.
As opposed to the integrated luminosity 
of the whole azimuthal angle, figure \ref{fig:fbaryon}
uses  the four-times luminosity of the filament 
and void directions.
The filament direction shows a systematically higher 
$K$-band luminosity than that in  void direction.
The integrated $K$-band luminosity
 profiles of these two directions become flatter from  $\sim$800-1000 kpc,
and the contribution of galaxies beyond 1000 kpc is small.
The integrated luminosity profile 
of the whole azimuthal angle  is  between those in  the filament 
and void directions except for the outermost region around 1200 kpc 
(Figure \ref{fig:fbaryon} and \ref{fig:extended source}).
This result is attributed to the luminosity of the
whole azimuthal angle being  affected by the filamentary structure around 1200 kpc 
in the south of the Hydra-A cluster.

We derived cumulative stellar mass fraction as,
\begin{equation}
f_{\rm star}(< r) = \frac{M_{\rm star}(< r)}{M_{\rm H.E.}(< r)}
\end{equation}
where, $M_{\rm star}(<r)$ is the total stellar mass within $r$,
derived from the total K-band luminosity within $r$, under the assumption that
the
stellar-mass-to-light ratio in the K-band is unity \citep{Nagino2009}.
At a given radius, the $f_{\rm star}(<r)$ in the filament direction
is higher by a factor of 2--3 than that in the void direction,
reflecting the concentration of galaxies  in the filament direction (Figure \ref{fig:fbaryon}).
Beyond $r_{500}$, $f_{\rm star}(<r)$ is smaller than $\sim$ 1\%,
and does not contribute to the baryon fraction.
We also derived the stellar mass fraction using  the whole azimuthal angle
using the NFW mass for the filament and void directions.
Although the stellar fraction continues to decrease with radius in the
filament and the void directions, 
the stellar fraction in whole azimuthal directions shows a minimum at $\sim r_{500}$,
and increases with radius by a factor of $\sim$1.5 from $r_{500}$ to $r_{100}$.

\begin{figure}[t]
     \FigureFile(80mm,60mm){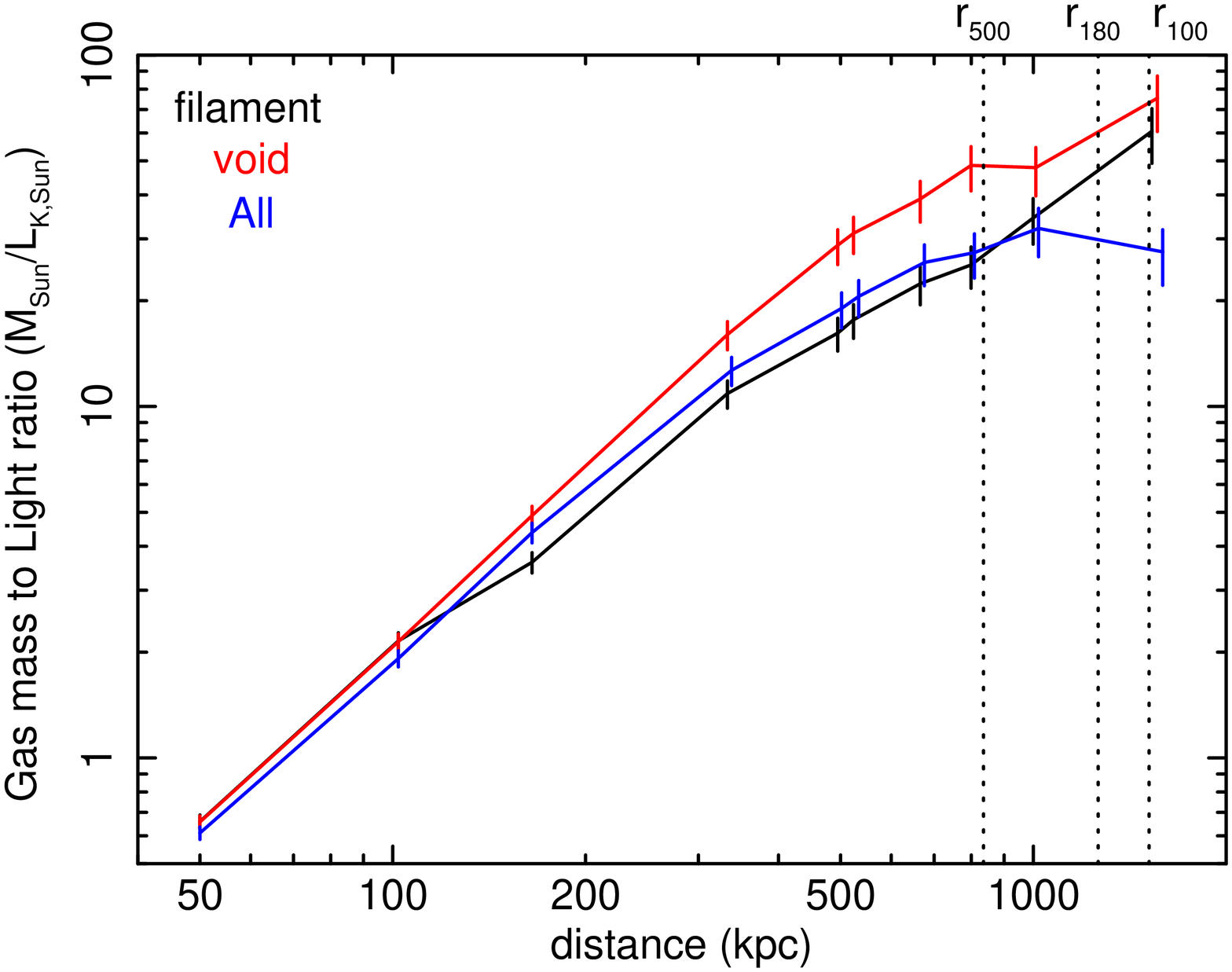}
\caption{Cumulative
 gas-mass-to-light ratio $M_{gas}/L_{K}$ in the filament (black) and void (red) directions
 and for the  whole azimuthal angle (blue).
}
\label{fig:gasratio}
\end{figure}

Using the integrated $K$-band luminosity profile, 
we calculated the {gas-mass-to-light ratio} in the 
filament and void directions (Figure \ref{fig:gasratio}).
The radial profile of the gas-mass-to-light ratio in the two directions increased
with radius out to $r_{100}$.
We also calculated the gas-mass-to-light ratio in the whole azimuthal angle, 
using  the  stellar luminosity in the whole azimuthal angle and
the gas mass in the void direction which agrees that for the whole
azimuthal angle derived by ROSAT \citep{ikebe97}.

\section{Discussions}

The Hydra A cluster is the first medium-sized cluster observed with Suzaku out
to the virial radius.
Suzaku observed the northern half of the cluster, and we derived
radial profiles of temperature, electron density, 
entropy,  hydrostatic mass, and  K-band stellar luminosity
 in the filament and void directions.
Temperature and entropy 
 carry important information about the thermal history of the ICM.
In subsection \ref{subsec:comparisonkT},
we compare these profiles of the Hydra A cluster and
other clusters observed with Suzaku and XMM
  to study their dependence on the system mass or
on  average ICM temperature.
Next, we discuss possible explanations for  entropy flattening and 
the higher gas-mass-to-hydrostatic-mass ratio 
at cluster outskirts.
Discrepancies between  electron and ion temperatures are discussed
in subsection \ref{subsec:thermodynamics},
deviations from hydrostatic equilibrium and the gas-clumping effect 
are discussed in subsections \ref{subsec:hydro} and
\ref{subsec:cm}.
We also derive IMLR of the Hydra A cluster out to the
virial radius in subsection  \ref{subsec:imlr}
to constrain the star formation history in the cluster.

\subsection{Comparison of temperature and entropy profiles with other systems}
\label{subsec:comparisonkT}

\begin{figure}[t]
\FigureFile(80mm, 60mm){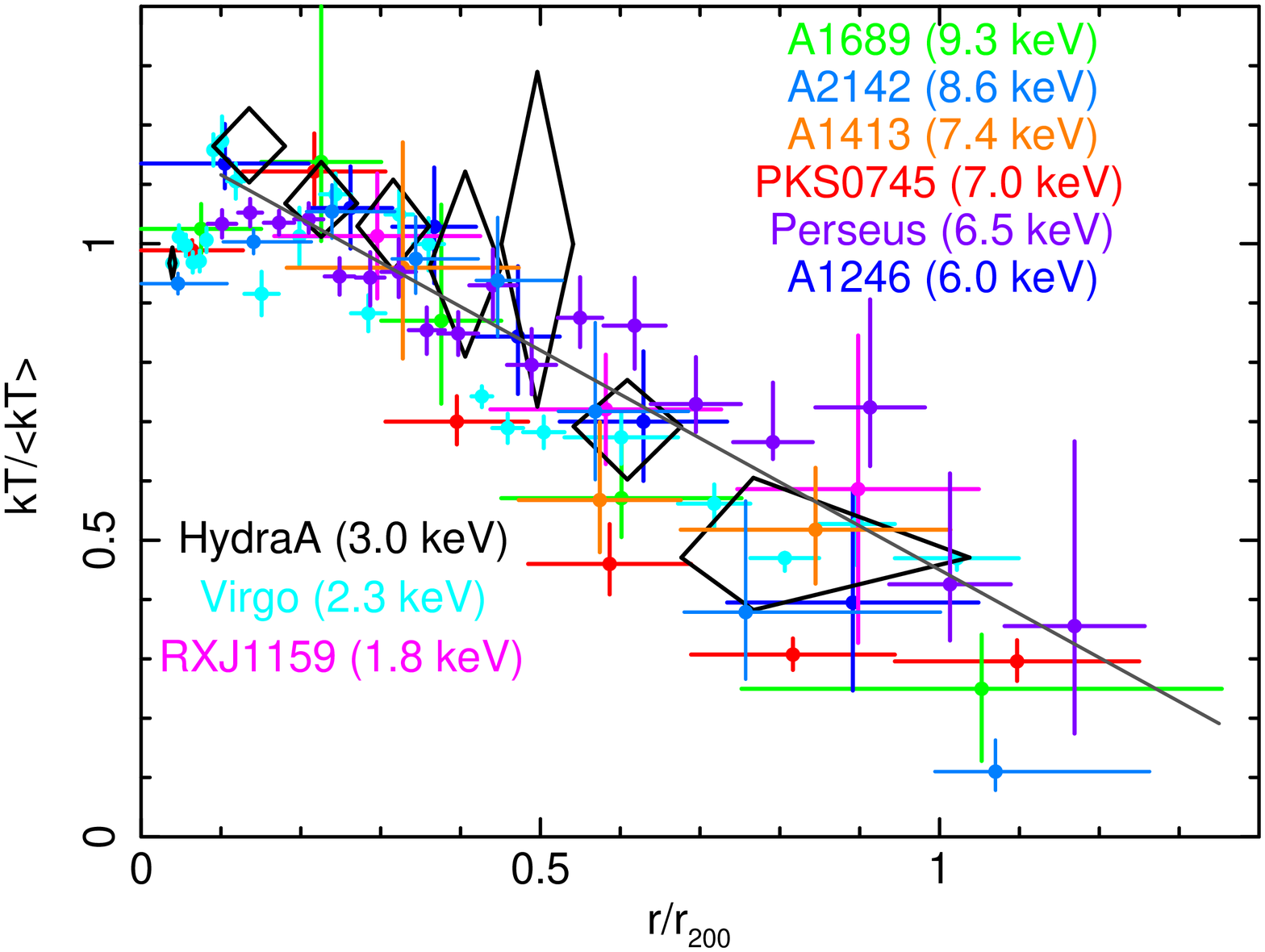}
\caption{Radial profiles of  temperature of clusters observed with Suzaku and XMM.
Here, the temperature profiles are scaled with the average ICM temperature, $<kT>$,
and are azimuthally averaged.
We plotted the radial profiles of
cluster of galaxies Abell 1689 \citep{kawaharada10}, Abell 2142 \citep{akamatsu11},
Abell 1413 \citep{hoshino10}, PKS-0745 \citep{george09}, the Perseus cluster \citep{simionescu11},
Abell 1246 (\cite{sato12}),  the Virgo cluster
\citep{urban11},  RXJ1159 \citep{humphrey11} and the Hydra A cluster (current study).
The scaling radius $r_{200}$ was calculated by \citet{henry09}.
The solid line represents the scaled temperature profile by \citet{pratt07}.
}
\label{fig:scalekt}
\end{figure}

Figure \ref{fig:scalekt} compares the temperature profiles of the Hydra A cluster
and several clusters observed with Suzaku, scaled with  average temperature, $<kT>$,
and $r_{200}$.
Here, 
$r_{200} = 2.77h^{-1}_{70}~\sqrt[]{\langle kT\rangle /10~{\rm keV}}/E(z)$ Mpc,
with $E(z)=(\Omega_{M}(1+z)^{3}+1-\Omega_{M})^{1/2}$ \citep{henry09}.
Then, $r_{200}$ of the Hydra A cluster is 1.48 Mpc.
Within 0.5$r_{200}$, the temperature profiles agree with the
scaled  temperature relationship  by \citet{pratt07}, which is
 derived from a sample of 15 clusters observed
with XMM out to $\sim r_{500}$.
In contrast, beyond 0.5$r_{200}$, the observed temperatures scatter by a
factor of two, and  tend to be lower than the XMM relationship.
In the scaled temperature profiles, there was no evidence
of any significant dependence on the average ICM
temperature, or system mass. 
Some variation in the temperature profiles should reflect
the azimuthal variation as observed in Abell 1689 \citep{kawaharada10} and Abell 1246
\citep{sato12}.

\begin{figure}[t]
\FigureFile(80mm, 60mm){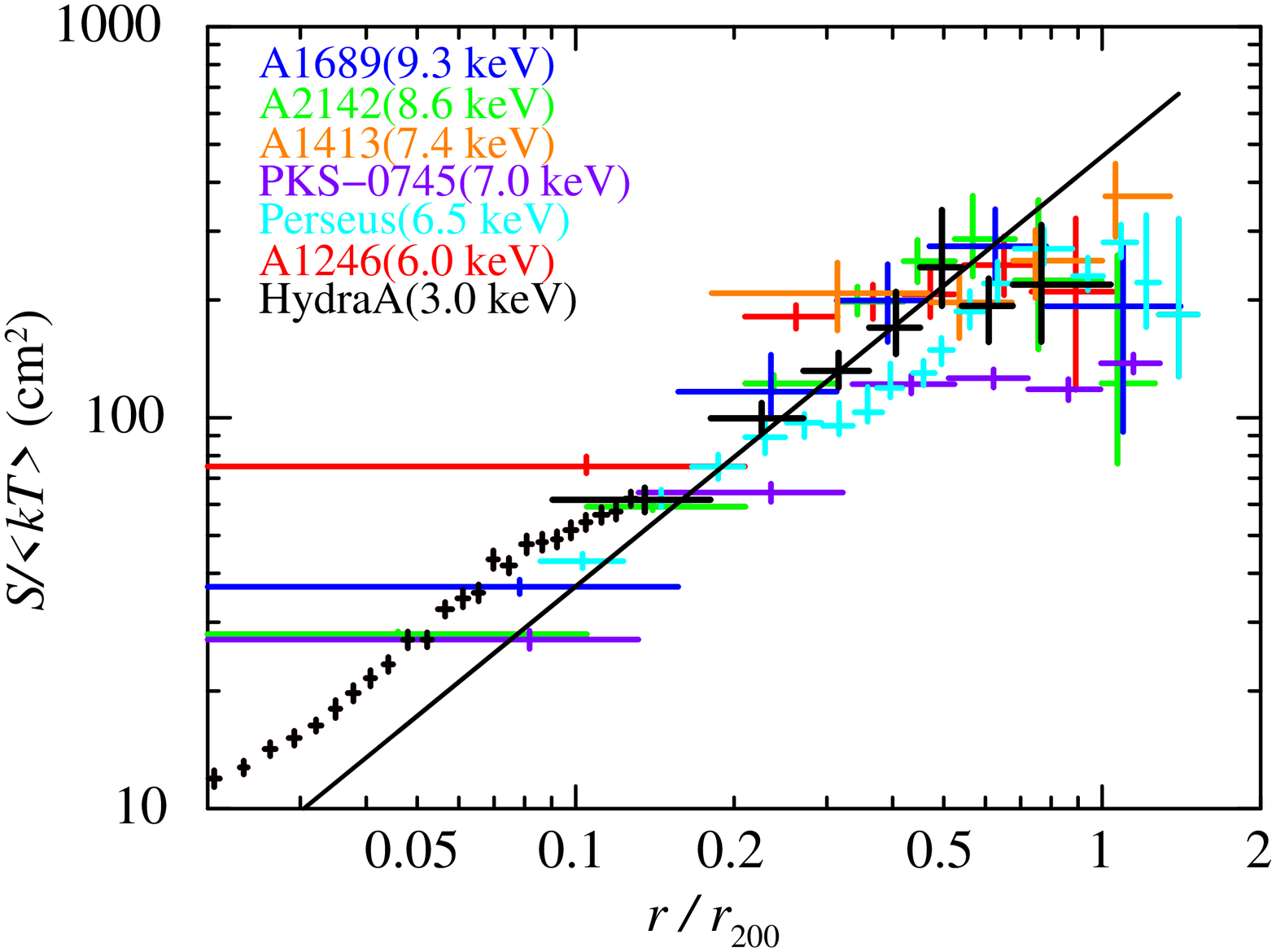}
\caption{Radial profiles of entropy of cluster observed with Suzaku and scaled:
with the average ICM temperature, $<kT>$, and are azimuthally averaged.
Data for
Abell 1689 \citep{kawaharada10} are plotted in dark blue,
for  Abell 2142 \citep{akamatsu11} in green, for
Abell 1413 \citep{hoshino10} in orange, for PKS-0745 \citep{george09} in purple,
for Perseus \citep{simionescu11} in light blue, for
Abell 1246 (\cite{sato12}) in red, and for Hydra A (current study) in black.
The scaling radius $r_{200}$ was calculated by \citet{henry09}.
The solid line shows the power-law profile with  slope of 1.1. 
The normalization of the power-law is scaled with our result for the Hydra A cluster.
}
\label{fig:scaledent}
\end{figure}

In figure \ref{fig:scaledent}, 
 we compare the entropy profile of the Hydra A cluster with
other clusters observed with Suzaku; these profiles are scaled with the
average ICM temperature and $r_{200}$ by \citet{henry09}.
Here, we  used the weighted average  of the radial profiles of
different directions of each cluster.
Contrary to the expected   $r^{1.1}$ relationship,
these profiles become flat
beyond 0.5$r_{200}$.
Except for PKS-0745 \citep{george09},
the scaled entropy profiles of these clusters 
were universal with ICM temperatures above 3 keV 
and did not depend on the ICM temperature.
For the PKS-0745 cluster, 
\citet{eckert11} determined that the ROSAT surface-brightness profile  
is statistically  inconsistent (7.7$\sigma$) with Suzaku results beyond $r_{200}$
and that the difference is likely explained by the existence of additional 
foreground components at the low Galactic latitude 
of the source. These components were not considered in the Suzaku background modeling.

\subsection{Ion-electron relaxation in the Hydra A cluster}
\label{subsec:thermodynamics}

One  interpretation for the  flattening of the entropy profile is that caused by
deviations in electron and ion temperatures \citep{hoshino10, akamatsu11}.
The time-scale for thermal equilibration between 
electrons and ions through Coulomb scattering is given by
\begin{equation}
t_{ei} \sim 0.14\left(\frac{n_{e}}{10^{-4}~\rm{cm^{-3}}}\right)^{-1}
\left(\frac{kT}{1.5~\rm{keV}}\right)^{3/2} \rm{Gyr}
\end{equation}
(\cite{takizawa1998a}; \cite{takizawa1998b}; \cite{takizawa1999}; \cite{akahori10}).
Considering this time-scale and the shock propagation speed, 
the radial length above which the electron temperature is significantly
lower than the ion temperature is proportional to the square of the ICM
temperature \citep{takizawa1999}. On the other hand, $r_{200}$ is proportional
to the square root of the ICM temperature. Therefore, some dependence 
on the mean ICM temperature is expected in the temperature and entropy
profiles.
However, there is no systematic dependence on the average ICM
temperature in the temperature and entropy profiles, as shown in figure
\ref{fig:scalekt} and figure \ref{fig:scaledent}. 
Furthermore, 
if the ICM temperature is underestimated and the actual 
entropy profile of the Hydra A cluster follows the $r^{1.1}$ relationship,
a flat ICM temperature profile is necessary, i.e., $r^{\sim 0}$ out to $r_{100}$,
considering the electron density profile of $r^{-(1.6\sim 1.7)}$ beyond 400 kpc.
However, the flat temperature profile contradicts to the 
results of numerical hydrodynamical simulations, 
where negative gradients of the temperature 
profiles outside cool core regions are naturally produced 
\citep{Borgani2002, burns10, nagai11}.
Therefore, for the Hydra A cluster, 
it is difficult to explain the flattening of the observed entropy profiles
by deviations of ion-electron temperatures.

\subsection{Deviations from hydrostatic equilibrium}
\label{subsec:hydro}

The remaining interpretation for the flattening of the entropy profiles at 
the cluster outskirt is
deviations from hydrostatic equilibrium, because when
the gas-clumping effect is significant, we also expect that
the ICM also  deviates from hydrostatic equilibrium.
Then, infalling matter may have retained some of its kinetic
energy in the form of bulk motions,
 and the thermal energy deficit in the ICM yields  lower entropy.
Furthermore, the underestimate of the gravitational mass and/or gas clumping 
leads to an overestimation of the baryon fraction.
Recent numerical simulations by \citet{vazza09} and \citet{nagai07} 
have shown that the kinetic energy of bulk motion carries $\sim$ 30\% of the total energy 
around the virial radius. 
With simulations, \citet{nagai11} showed that beyond $r_{200}$, gas clumping leads 
to an overestimation  of the observed gas density and causes flattening of the entropy profile.
If the hydrostatic mass was underestimated by 130\% in the filament direction and 100\% 
in the void direction, the baryon fractions at $r_{100}$ would have the same value as that of WMAP7, although the observed gas fraction calculated from the
hydrostatic mass in the void direction is consistent with the WMAP7 result within error bars.

The distributions of the ICM and galaxies in the Hydra A 
cluster elongate in  the filament direction.
If the ICM is under hydrostatic equilibrium in a non-spherical dark matter
halo, the X-ray emission and dark matter halo elongate in the same direction.
Numerical simulations by \citet{burns10} determined that 
gas, galaxies, and dark matter continue in the filament direction due to the 
accretion of gas and subclusters.
Then, it is reasonable that the dark matter halo also elongates in the filament direction.
However, the hydrostatic mass in the void and filament directions of the Hydra A cluster
are in good agreement.
These results indicate that
deviations from  hydrostatic equilibrium
should be more significant in the filament direction than  in the void direction.

The  clumping effect  will  also be  higher in the filament direction, since
 smaller systems are thought to be accreted from this direction.
For the Hydra A cluster, we were not able to detect clump candidates  with Suzaku.
The spectra of detected point sources are consistent with those of
background active galactic nuclei.
Considering the luminosities of detected point sources,
the luminosity of clumps should be smaller than
a few 10$^{41}\ {\rm erg\ s^{-1}}$, which
 corresponds to cores of groups of galaxies.
Therefore, if clumps are significant, their scales should be smaller than
those of small groups of galaxies.

\subsection{The concentration-mass relationship}
\label{subsec:cm}

\begin{figure}[t]
\FigureFile(80mm, 60mm){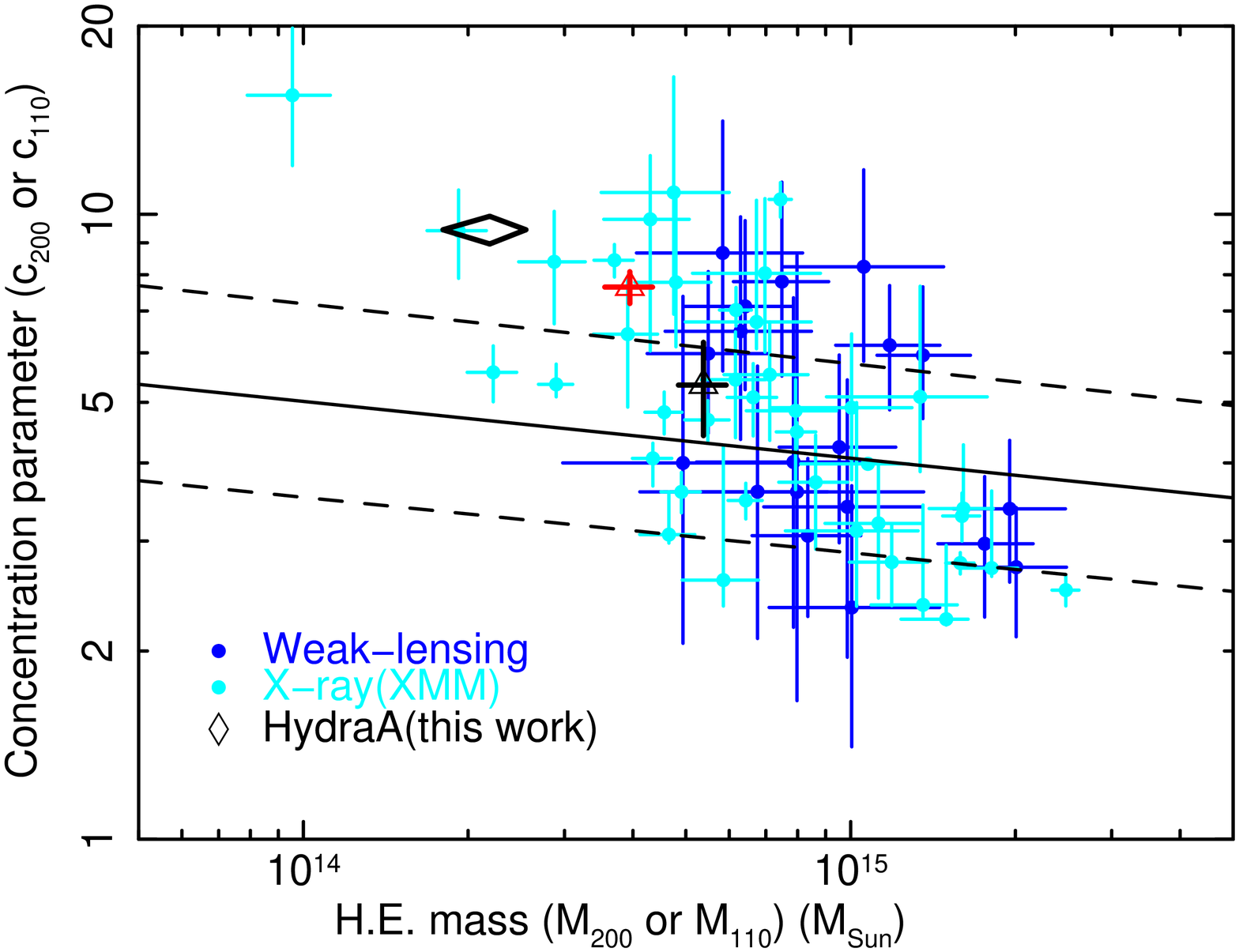}
\caption{
 Observed distribution of  concentration parameters
as a function of cluster masses. The mass of weak lensing (blue) and X-ray (light blue)
were defined by \citet{okabe10} and \citet{ettori10};\citet{ettori11}, respectively. 
\citet{okabe10} used $M_{110}$ and $c_{110}$, while \citet{ettori10};\citet{ettori11}
used $M_{200}$ and $c_{200}$.
We considered $M_{100}$ for Hydra A.
The solid line indicates the median relationship
determined through CDM simulations for the WMAP5 cosmological model, while 
the region enclosed within the dashed lines corresponds to the range of
$\sigma({\rm log_{10}}c)=0.1$, within which most of simulated clusters are distributed
\citep{duffy08}.
The large diamond shows the observed mass profile of Hydra A (current work)
assuming hydrostatic equilibrium.
Black and red open triangles correspond to cases in which 
the $f_{\rm gas}(<r)$ value of Hydra A (current work)
is the same as the cosmic mean baryon fraction
\citep{komatsu11} in the filament and the void directions, respectively.
}
\label{fig:lensmass}
\end{figure}

From numerical simulations, 
a weak variation in the concentration is expected from low-mass to high-mass clusters,
reflecting differences in the formation epochs of low-mass and high-mass halos
(\cite{navarro1997}; \cite{bullock01}; \cite{dolag04}).
\citet{pointecouteau05}, \citet{ettori11} and \citet{okabe10} 
investigated the relationship between the concentration 
parameter and the cluster mass, or the concentration-mass ($c-M$) relationship.
The results are compatible with the intrinsic
dispersion of theoretical predictions.
Figure \ref{fig:lensmass} compares the $c-M$ relationship of
the Hydra A cluster and other clusters observed with weak-lensing \citep{okabe10},
and clusters observed with XMM by \citet{ettori11}.
An assumption of cluster dynamical state or 
hydrostatic equilibrium is not required in the weak-lensing results.
The distribution of the $c-M$ relationship of these clusters
 agrees well with the relation expected from the numerical simulations
by \citet{duffy08},
although 
\citet{ettori11} found that relatively low-mass systems 
tend to have higher concentration parameters.

The hydrostatic mass in the Hydra A cluster 
of the filament and void directions are represented with the same NFW mass model
with   $c_{\rm vir}=9-10$ and $M_{\rm vir}=(2.2\pm0.4)$ $\times 10^{14}~M_{\odot}$.  
As shown in Figure \ref{fig:lensmass}, this  $c_{\rm vir}-M_{\rm vir}$
relationship  of the Hydra A cluster was outside $\sigma(\log_{10}c)=0.1$ within
which most  clusters from numerical simulations are distributed \citet{duffy08}.
If the hydrostatic mass was underestimated and
 the baryon fraction using the gravitational mass beyond $r_{500}$
of the Hydra A cluster is the same as the cosmic mean baryon fraction
in the two observed directions, the fits with
 the NFW model yielded smaller values of $c_{\rm vir}$ and higher
values $M_{\rm vir}$ as shown in Figure \ref{fig:lensmass}.
In this case,
the derived value of $M_{\rm vir}$ in the filament direction became higher than
that in the void direction.
This result also indicates that the ICM in the outskirts of Hydra A
deviates from hydrostatic equilibrium, and 
the X-ray emission and dark matter halo elongate in the same direction.

\subsection{Metal-Mass-to-Light Ratios out to the virial radius}
\label{subsec:imlr}

\begin{figure}[t]
     \FigureFile(80mm,60mm){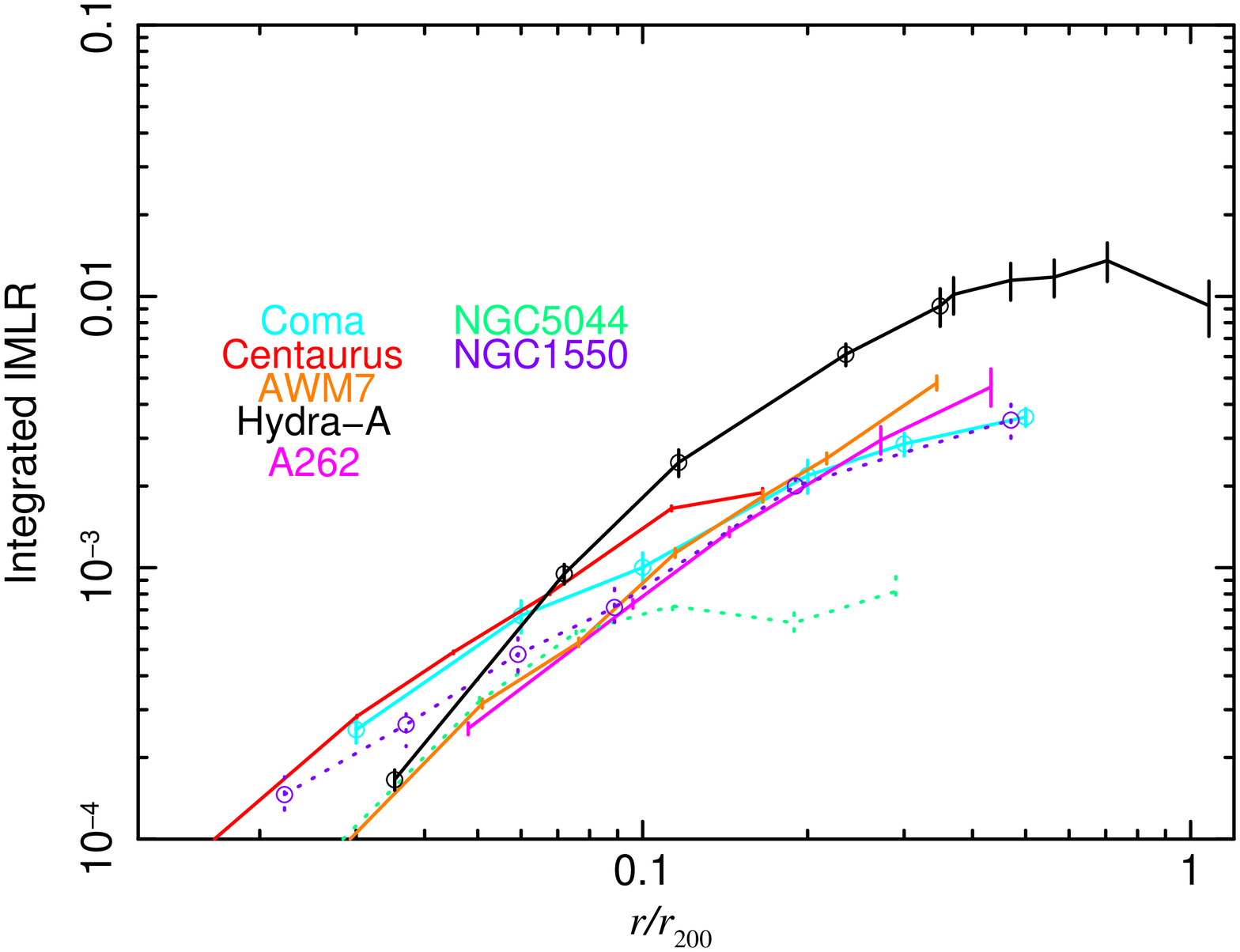}     
\caption{
Comparison of iron-mass-to-light-ratio for other clusters. 
The radius is scaled
with the characteristic radius $r_{200}$ \citep{henry09}.
 For comparison, we plotted the radial profile of cluster of
galaxies AWM 7 (\cite{sato08}, Sato et al. in preparation), Abell 262 \citep{sato09b}, 
the Centaurus cluster \citep{sakuma11} and the Coma cluster
\citep{matsushita2011b}
and galaxy groups NGC 5044 \citep{komiyama09} and NGC 1550 \citep{sato10}. 
Open circles are the result from XMM-Newton satellite, 
and others are Suzaku satellite. }
\label{fig:imlr}
\end{figure}
     
We derived integrated IMLR,  mass-to-light ratio for Fe,
 out to the virial radius  for the first time.
Here, we show the IMLR profile for the whole azimuthal angle to compare
other clusters observed with Suzaku.
Within 0.3 $r_{200}$ and beyond 0.3 $r_{200}$,
we used the Fe abundances  obtained by \citet{matsushita11} and this work, respectively.
Here, within 8$'$--15$'$, or 0.35--0.65 $r_{200}$, we used the weighted average of the
Fe abundance at the three radial bins in this region.
The error bars of the mass-to-light ratio  include the errors in both the abundance 
and  the K-band luminosity caused by the uncertainty in the 
contribution of fainter galaxies below the 2MASS threshold
  (see section \ref{subsec:imlr} for details).
Figure \ref{fig:imlr} shows the cumulative IMLR profile of the Hydra A cluster. 
The IMLR profile increases with radius within 0.4--0.5 $r_{200}$, and
becomes flatter beyond 0.5 $r_{200}$ with the value of $\sim 0.01~ M_\odot/L_{K,\odot}$,
which reflects the flattening of the gas-mass-to-light ratio beyond $r_{500}$.
For the void and filament directions,   the IMLR profiles should increase with radius
beyond $r_{500}$.

At $r_{200}$, the cumulative IMLR was 0.007--0.01 $M_\odot/L_{K,\odot}$.
A previously reported 
theoretical model predicts that the oxygen-mass-to-light ratio (OMLR)  
 of a cluster is a sensitive function of
the slope of the initial mass function (IMF) \citep{Renzini2005}.
Here, the oxygen mass is a sum of that trapped in stars and that in the ICM.
By adopting a Salpeter IMF with a slope of 2.35, the expected value of
the OMLR is $\sim0.1~ M_\odot/L_{B,\odot}$.
In contrast, a top-heavy IMF with a slope of 1.35 overproduces metals more than 
that with a factor of 20.
For the Hydra A cluster, by
adopting the ratio of stellar-mass-to-light ratio 
 4--6 in  the B-band and K-band  \citep{Nagino2009, matsushita2011b},
the IMLR in the B-band becomes 0.03--0.06  $M_\odot/L_{B,\odot}$ and 
the cumulative  OMLR within the virial radius becomes
$\sim $0.1--0.3 $\left(A_{\rm O}/A_{\rm Fe}\right) M_\odot/L_{B,\odot}$.
Here, $A_{\rm O}$ and $A_{\rm Fe}$ are abundances of O and Fe in solar units,
respectively.
Beyond $r_{500}$ of the Hydra A cluster, 
the integrated stellar mass is only several percent
of the integrated gas mass.
Thus, even by considering the difference of stellar metallicity and the abundance of the ICM,
most of the metals in the cluster are in the ICM.
Therefore, for cases with solar O/Fe ratio, the OMLR
is similar to the expectations reported by \citet{Renzini2005},
assuming a Salpeter IMF in the Hydra A cluster. 
In contrast, by adopting the O/Fe ratio from a nucleosynthesis model
of SN II with metallicity $=0.02$,  as reported by \citet{Nomoto2006},
the OMLR within the virial radius becomes 0.3--1.0 $M_\odot/L_{B,\odot}$
in the Hydra A cluster.
This value is larger than the expectation from the Salpeter IMF,
and is consistent with a flatter IMF slope.

In Figure \ref{fig:imlr}, the cumulative IMLR profile
 of the Hydra A cluster was compared with those of other groups and clusters,
including Coma (8 keV; \cite{matsushita2011b}),
Centaurus (4 keV; \cite{sakuma11}),
and AWM 7 clusters (3.6 keV; \cite{sato08}), 
as well as NGC 1550 (1.2 keV; \cite{sato10}) and NGC 5044 groups (1.0 keV; \cite{komiyama09}).
Within $0.5r_{200}$,
the cumulative IMLR profile of the Hydra A cluster increased with
radius by a factor of two higher than for other systems.
No systematic dependence on the ICM temperature was evident for clusters
with temperatures higher than 2 keV,
although beyond 0.1$r_{200}$, a group of galaxies shows a significantly smaller IMLR 

The stellar and gas-mass fractions within $r_{500}$ depend on the total
system mass \citep{lin03, lin04, Vik06, Sun09, Gio09}.
These studies determined that within $r_{500}$ the
 stellar-to-total-mass ratios  of the groups are much 
larger than those in the clusters, whereas the gas-mass fraction increases with
the system mass.
The observed higher stellar mass fraction and the lower gas mass fraction
 within $r_{500}$ in poor systems
are occasionally interpreted as proof  that the
 star formation efficiency depends on the system mass.
However, 
as shown in Figure \ref{fig:gasratio}, 
\color{black} the gas is more extended than the stars in a cluster within $r_{500}$.
The  gas density profiles in the central regions of groups and poor
clusters were observed  to be shallower than those in the self-similar model,
and the relative entropy level was correspondingly higher than that
 in rich clusters \citep{Ponman99, ponman03, Sun09}.
Then, the difference in the ratio of gas-mass-to-stellar-mass may reflect
differences in distributions of gas and stars, which in turn reflects the history of 
energy injection from galaxies to the ICM.
To study the fractions of stars and gas in clusters of galaxies, and
its dependence on the system mass, 
we need measurements of gas and stellar mass beyond $r_{500}$ of  other clusters.

\section{Summary and Conclusion}

We presented the results of the Suzaku observation of the Hydra A galaxy cluster, 
which is the first medium-sized cluster 
(temperature $\sim$ 3~keV) observed with Suzaku out to the virial radius. Two observations were 
conducted, north-west and north-east offsets,
which continue in a filament direction and a void direction
 of the large-scale structure of the Universe,
respectively. We investigated possible azimuthal variations 
in temperature, electron density, entropy  and total mass profiles. 
Our analysis revealed that distributions of X-ray emission and galaxies
elongate in the filament direction.
The entropy profiles become flatter beyond $r_{500}$ in contrast 
to the $r^{1.1}$ relationship expected from the shock heating model of the ICM. 
The entropy profiles are universal in clusters observed with Suzaku
when  scaled with the averaged temperature of each cluster.  
The hydrostatic masses in the two directions agree and
the NFW universal matter profile represents the hydrostatic mass distribution obtained
 up to $1.8~r_{500}$,  with 
$c_{\rm vir}=9-10$ and $M_{\rm vir}=(2.2\pm0.4)\times 10^{14}M_{\odot}$.
The gas fraction, $M_{\rm gas} (<r)/M_{\rm H.E.} (<r)$ significantly exceeds
the cosmic mean baryon fraction of WMAP7 beyond $\sim r_{500}$.
The flattening of the entropy profile and higher gas fraction contradict
expectations based on numerical simulations.
An underestimate of gas temperature due to the discrepancy between ion and electron 
temperatures gives higher entropy and smaller gas fraction at cluster outskirts.
However, a flat or increasing temperature profile is required.
If bulk motions caused by infalling matter from filaments of clusters
are higher than that of numerical simulations, and if the ICM deviates from the
hydrostatic equilibrium, the entropy becomes smaller and the gas fraction is overestimated.
In addition, we obtained IMLR up to the virial radius for the first time,
and compared our results with other clusters and groups.
The IMLR profile is consistent within 0.5 $r_{200}$ with other clusters,
and become flatter from 0.5 $r_{200}$ to $r_{200}$.

\end{document}